\documentclass[11pt]{article}
\usepackage{amssymb}
\newlength{\Mylen}
\title{Decomposing Quantum Fields on Branes}

\author{Marco Bertola$^1$, Jacques Bros$^3$,
Vittorio Gorini$^{2}$, Ugo Moschella$^{2}$
and { {Richard Schaeffer}}$^{3}$
\\{\small $^{1}$ C.R.M.,  Universit\'e\ de Montr\'eal, C.P. 6128
Succ. Centre-Ville, Montr\'eal (Qu\'ebec), H3C 3J7. }\\
{\small$^{2}$Istituto di Scienze Matematiche Fisiche e Chimiche, Via
Lucini 3, 22100 Como,}
\\ {\small{and INFN sez. di Milano, Italy}.}
\\{\small $^{3}$ Service de Physique Th\'eorique, C.E. Saclay,
91191 Gif-sur-Yvette, France.}}

\def\X{{\cal X}}
\def\Y{{\cal Y}}
\def\K{{\varkappa}}
\def\PHI{{\widehat {\Phi}}}

\def\Varphi{{\widehat\varphi}}

\newcommand{\be}{\begin{equation}}
\newcommand{\ee}{\end{equation}}
\newcommand{\ba}{\begin{eqnarray}}
\newcommand{\ea}{\end{eqnarray}}
\newcommand{\pa}{\partial}
\def\bra#1{\le\langle{#1}|}
\def\vrul{\rule[20pt]{0pt}{0pt}}
\def\ket#1{{#1}\ri\rangle}
\def\bra#1{\le\langle{#1},}
\def\bea{\begin{eqnarray}}
\def\eea{\end{eqnarray}}
\def\le{\left}
\def\ri{\right}

\newtheorem{theo}{\sc Theorem}
\newtheorem{defi}{\sc Definition}
\def\bd{\begin{defi}}
\def\bt{\begin{theo}}
\def\et{\end{theo}}
\def\ed{\end{defi}}

\def\l{\lambda}

\def\R{{\mathbb R}}

\def\M{{\cal M}}
\def\a{{\rm a}}
\def\b{{\rm b}}
\def\kk{{\rm k}}
\def\ll{{\rm l}}

\def\bes{$$}
\def\ees{$$}
\def\beas{\begin{eqnarray*}}
\def\eeas{\end{eqnarray*}}

\def\o{\omega}

\begin{document}
 \maketitle


\abstract{We provide a method
to decompose the two-point function of a
quantum field on a warped manifold in terms of fields living on a
lower-dimensional manifold.
We discuss  explicit applications
to Minkowski,  de Sitter and anti-de Sitter quantum
field theories. This decomposition presents a remarkable analogy with
the holography principle, in the sense that physics in $d+1$
dimensions may be encoded into the physics in one dimension
less. Moreover in a context {\em \`a\ la} Randall--Sundrum, the method
outlined here allows a mechanism of generation of mass-spectra in the
3-brane (or more generally a $d-1$-brane).}
\section{Introduction}
The idea of dimensional reduction in Quantum Field Theory is very
old, dating back to the Kaluza--Klein theory. The motivation for
considering theories in a larger ambient space is the hope to
simplify or unify certain aspects of the lower dimensional
theory. Indeed one expects that the extra degrees of freedom of
the field in  the ambient space survive somehow encoded in the
restricted theory.\par The basic ingredient of such an approach
is to embed the spacetime of interest into a larger manifold and
then consider an extension of the field to this ambient space in
order to read off the properties of the original field into the
(hopefully easier) formulation of the theory in the ambient
manifold.

 To make an example it is known that a QFT on the de
Sitter spacetime manifests thermal properties to an inertial
observer
\cite{Gibbons:1977mu,Bros:1996js,Bros:1994dn,brosepsteinmoschella}:
this is a kind of Hawking effect adapted to the present geometry.
However, if we regard the de Sitter manifold as a submanifold of
an ambient Minkowski (hence flat) manifold, what is an inertial
observer in de Sitter becomes a uniformly accelerated observer in
the ambient flat spacetime. This allows us  to regard the de
Sitter thermal effect as a Unruh effect in the higher dimensional
flat spacetime \cite{bertola,deser}.

Moreover the general status of field theory in  flat spacetime is
well established \cite{Streater,itz}, which is not true for
generic curved spacetimes \cite{birrel,wald}: the possibility of
embedding the de Sitter manifold in the ambient Minkowski space
allows one to formulate a sort of Wightman axiomatic framework
for de Sitter spacetime, as if ``geometrically'' inherited from
the existing axioms of the Minkowskian case
\cite{brosepsteinmoschella,Bertola:1999ga}. In perspective this
approach seems to be quite promising.

In the present work we address this kind of problems in the
rather general framework of ``warped manifolds'': these are
obtained by a topological product of manifolds, a ``base'' and a
``fiber'' or ``leaf'' (or ``brane'').

As a pseudo--Riemannian manifold the metric is obtained by warping
the metric of the fiber  by a scalar function $\o$ depending on
the point of the base.

Quite recently \cite{Randall:1999, Randall:2000} this sort of
warped manifolds have made their appearance in the context of the
``hierarchy problem''. There, the study is carried out in the
case in which the five dimensional background metric is made up
by gluing together two slices of the five dimensional anti-de
Sitter spacetime $AdS_5$.

The purpose of  the present paper  is to deal with a general
situation in which the extra dimensions are warping the ``brane''
by an arbitrary warp factor $\o$, which ultimately might be
considered as a further degree of freedom of the full theory.

Particularly relevant is the case of  only one extra dimension:
under that hypothesis we will be led to the study of an auxiliary
Schr\"odinger operator $\mathcal L$ in the extra dimension. Then
we will  prove that any (free) field $\PHI$ moving in the
background geometry will be seen by an observer in the 3-brane as
a bunch of fields $\Varphi_\l$ of different masses $m^2=\l$: the
spectrum of the allowed masses is dictated by the spectrum of the
Schr\"odinger operator $\mathcal L$.

As a matter of fact the treatment does not rely in any step on
the dimensionality of the embedded brane, hence we can replace
the 3-brane by any $d-1$-brane.

As we will see, warped products occur in quite a number of
relevant examples, the first to be mentioned being the previous
example of de Sitter and Minkowski. Indeed we can regard (a
suitable open subset of) the flat spacetime as a warped manifold
where the $d$-branes are de Sitter manifolds fibered on the half
line parameterizing their curvature radius. Other examples will
involve foliations of de Sitter manifolds by lower dimensional de
Sitter ones, or anti-de Sitter foliated by Minkowski manifolds.

The geometric structure of these warped manifolds enables us to
formulate precise correspondences between scalar Klein--Gordon
fields propagating in the ambient spacetime and the restriction
of them to a fixed fiber. In particular we show that under
suitable assumptions and in all the examples the restricted field
is a generalized free field admitting a generalized
K\"allen--Lehmann decomposition \cite{kallen} in terms of
Klein--Gordon fields propagating along the $d-1$-brane.\par\vskip
3pt
The plan of the paper is the following: in Sections \ref{foreplay}
and 1.2, we expose some elementary facts about the canonical Klein--Gordon
theory in the flat spacetime, using it as a toy--model to introduce  
the ideas developed in the following. \\
In Section \ref{warped} we provide the general framework of
Klein--Gordon QFT on warped manifolds.\\
In Section 3, we enrich the previous framework by imposing
appropriate ``consistency conditions'' between the geometries of
the ``bulk'' and of the ``brane''. By using the latter, we
provide in Section 4 a complete treatment of the aforementioned
examples, namely of the correspondences de Sitter--Minkowski (Sec.
\ref{app1}), de Sitter--de Sitter (Sec. \ref{dS2dS}),
Unruh--Minkowski (Sec. \ref{app2}) and Minkowski--Anti de Sitter
(Sec. \ref{app3}). This latter application has relevance in the
aforementioned context of the hierarchy problem as well as in the
AdS/CFT correspondence \cite{Maldacena:1997re} as it has been
pointed out in \cite{BBMS:1999}.

\subsection{Canonical Klein--Gordon field theory}
\label{foreplay} We begin with a quick review of ordinary
Klein--Gordon theory in Minkowskian spacetime in order to
illustrate the idea of the paper.\par Let us consider the
$(d+1)$-dimensional Minkowski spacetime ${\mathbb M}^{d+1}$ with
inertial coordinates $(X^0,X^1,$ $\ldots,X^d)$ and  metric \be
ds^2_{d+1} = {dX^0}^2 -{dX^1}^2 - \ldots - {dX^d}^2\ . \ee Let
$\PHI$ be a Klein--Gordon quantum field of mass $M$ in the
Wightman vacuum: \be \left(\square_{d+1} + M^2 \right)\PHI = 0.
\label{kkgg} \ee The field $\PHI$ can be represented in terms of
standard creation and annihilation  operators and one deduces the
momentum space (Fourier) representation for the two--points
correlation function of $\PHI$: \be W^{(d+1)}_M(X,X')= \langle
\Omega, \PHI(X)\PHI(X')\, \Omega\rangle = \frac{1}{(2\pi)^d}\int
_{{\mathbb R}^{d+1}} e^{-i  P\cdot (X-X')} 
{\Theta}(P_0)\delta(P^2-M^2) \ {{\rm d}^{d+1}P}  , \label{01} \ee where
$\Omega$ is the standard Wightman vacuum state and ${\Theta}$
denotes the Heaviside function. Let us  consider now the
restriction of the two-point function $W^{(d+1)}_m(X,X')$ to the
hyperplane
 $\Y=\{ X \in {\mathbb M}^{d+1}: X^d = x= const\}$.
$\Y$ inherits its metric  from the
ambient Minkowski spacetime and can be identified with a
$d$-dimensional Minkowski spacetime.

Since the restriction of $W^{(d+1)}_M(X,X')$  defines   an
acceptable two-point function (and therefore a generalized free
field) on $\Y\simeq {\mathbb M}^d$, it is  possible to decompose
it into elementary components, namely to construct its
K\"allen-Lehmann representation. This is particularly simple,
since  the representation (\ref{01})  can be rewritten as follows:
\begin{eqnarray}
 && W^{(d+1)}_M(X,X') =
\frac{1}{2\pi}\int_{M^2}^\infty
  \frac{\cos \left[\sqrt{\mu^2-M^2}(
x-x')\right]}{\sqrt{\mu^2-M^2}} W^{(d)}_\mu(y,y')\,{ {\rm d}(\mu^2)},
\label{hip}
\end{eqnarray}
where we have  introduced the   notations  ${y}=
(y^0\!,y^1\!,\ldots$,$y^{d-1}) $
with \ $y^0=X^0,...,$ $ y^{d-1}\!\! =X^{d-1}$, $x=X^d$ and
$\mu= {P^d}$. It follows that
\be
  W(y,y') = W^{(d+1)}_M(X,X')_{\downharpoonright_{\Y\times \Y}}
=\frac{1}{2\pi} \int_{\mu^2=M^2}^\infty  \!\!\! W^{(d)}_\mu(y,y')
 \frac{ {\rm d}(\mu^2) }{\sqrt{\mu^2-M^2}}.
\ee
This formula is a particular instance
of the well--known K\"allen--Lehmann decomposition.

\subsection{Spectral analysis}

Let us review this elementary example to single out its key
points. First of all the Minkowski manifold ${\mathbb M}^{d+1}$
can be written as the Cartesian product $ {\mathbb M}^{d+1} =
{\mathbb R}\times \Y$. Correspondingly the metric splits into
two  parts $ds^2_{d+1} =  - {dx}^2+{ds}^2_{d}$. This splitting
allows separation of variables in the Klein--Gordon Eq.
(\ref{kkgg}), giving rise to the following pair of equations for
the modes: \ba && \left(\square_{d} + \lambda\right)\varphi(y) =
0,
\\
&& \left( -\frac{\pa^2}{\pa {x}^2}+ M^2     \right) \theta(x) =
\lambda   \theta(x). \label{gti} \ea Now we can think of Eq.
(\ref{gti}) as a spectral problem in the Hilbert space
$L^2(\mathbb R)$, and look for a complete set of eigenfunctions
for the self-adjoint positive operator $\left(- \frac{\pa^2}{\pa
{x}^2} +  M^2 \right)$, which is a Schr\"odinger operator with
constant potential. It is useful to adopt  real--valued
eigenfunctions: \be \theta^1_\lambda(x) =
\frac{1}{\sqrt{2\pi\sqrt{\lambda - M^2}}} \cos \le(x
\sqrt{\lambda-M^2}\ri)\, , \;\;\;\;\;\; \theta^2_\lambda(x) =
\frac{1}{\sqrt{2\pi\sqrt{\lambda - M^2}}} \sin \le(x
\sqrt{\lambda-M^2}\ri), \ee with  $\lambda\geq M^2$. This set of
eigenfunctions is orthonormal and complete: \bea && \int_\R {\rm
d}x\, \theta^{(i)}_\lambda(x)\theta^{(j)}_{\lambda'}(x)=
\delta_{ij}\delta(\lambda-\lambda'),\label{comp1}\\
&& \sum_{i=1}^2\int_{M^2}^\infty  {\rm d}\lambda\,
\theta^{(i)}_\lambda(x)\theta^{(i)}_{\lambda}(x')=\delta(x-x').\label{comp2}
\eea
Let us introduce the following ``formal quantum fields'':

\be
\Varphi_{\lambda }^{(i)}(y)=\int_{\mathbb R }
\PHI(X) \theta_{\lambda}^{(i)}(x) {\rm d}x.
\label{huu}
\ee

We have used this terminology  (``formal'') to
indicate that in the Hilbert space of the Klein--Gordon field $\PHI(X)$ they are
operator-valued distributions not only with respect to $y$ (as usual), but
{\it also with respect to the mass parameter} $\lambda$, as it will appear below explicitly in the
expression of the two-point functions of these fields.

Eqs. (\ref{kkgg}) and (\ref{huu}) yield (in the sense of
distributions in the joint variables $(y, \lambda)$): \be
\left(\square_d + \lambda\right)\Varphi_{\lambda }^{(i)}(y) = 0.
\ee Furthermore these fields commute with each other for
different values of the parameter $\lambda$; actually, as it
results from Eqs. (\ref{hip}) and (\ref{huu}), their mutual
two-point correlation functions have the following expression: \be
W^{ij}_{\lambda,\lambda'}(y,y') = \langle \Omega,
\Varphi_{\lambda }^{(i)}(y) \Varphi_{\lambda' }^{(j)}(y')\,
\Omega\rangle = \delta_{ij} \delta(\lambda-\lambda')
W^{(d)}_{\sqrt{\lambda}}(y,y') {\Theta}(\lambda-M^2). \label{111}
\ee By inverting Eq. (\ref{huu}) we obtain \be \PHI
(X)=\sum_{i=1}^2 \int_{M^2}^\infty \Varphi_{\lambda }^{(i)}(y)
\theta_{\lambda}^{(i)}(x) {\rm d}\lambda . \label{112} \ee The
previous inversion formula has been obtained by means of the
completeness relation (\ref{comp2}). It is worthwhile to stress
that  this is  justified in the present case  because the field
$\PHI$ is a tempered operator--valued distribution and the theory
of inversion of Fourier transform extends to  tempered
distributions  \cite{Reed} (namely we are making a
Fourier--transform of  a tempered operator--valued distribution
w.r.t. the variable $x$  and taking its
inversion).\\
A straightforward computation using Eq. (\ref{111}) and Eq.
(\ref{112}) shows that \ba &W^{(d+1)}_M(X,X') & =
\sum_{i=1}^2\int_{M^2}^\infty {\rm d}\lambda
\theta_\lambda^{(i)}(x) \theta_{\lambda}^{(i)}(x')
W^{(d)}_{\sqrt{\lambda}}(y,y'), \ea formula which agrees with Eq.
(\ref{hip}) which was worked out directly.\par By { restriction}
of the  field $\PHI$ to the branes  of constant coordinate $x=x'$
we obtain \bes \sum_{i=1}^2\int_{M^2}^\infty {\rm d}\lambda
\le|\theta_\lambda^{(i)}(x)\ri|^2 W^{(d)}_{\sqrt{\lambda}}(y,y')
= \int_{M^2}^\infty \frac {{\rm
d}\lambda}{2\pi\sqrt{\lambda-M^2}} W_{\sqrt{\lambda}}^{(d)}(y,y')
. \ees
The spectral weight $\sum_{i=1}^2
\le|\theta_\lambda^{(i)}(x)\ri|^2= \frac
{1}{2\pi\sqrt{\lambda-M^2}}$ is the
{\em density of states per unit spectrum per unit volume}
 of the self--adjoint operator  $H =-\frac
{\pa^2}{\pa{x}^2} +M^2$.\par
We are going to extend this picture to more general manifolds in
the following sections.
\section{Klein--Gordon fields on warped manifolds: an expansion
formula.}
\label{warped}
The previous example suggests the following general structure.
Let $\left(\X,{ {^{(\X)}\!g}}\right)$
 be a Riemannian  manifold,
$\left(\Y,{^{(\Y)}\!g}\right)$  a $d$-dimensional pseudo--Riemannian (Lorentzian) manifold
and $\omega \in C^\infty\left(\X,\R^+\right)$
be a smooth positive function.
Define  $\M = \X\times \Y$  as a topological manifold.
The metric on $\M$ is defined by
\be
ds^2={g}_{\mu\nu} dX^{\mu}dX^{\nu} = ds^2_{\X}+\omega^2(x)ds^2_{\Y}\ ,
\ee
where
\be
ds^2_\X = { {^{(\X)}\!g}}_{\a\b} dx^{\a}dx^{\b},\,\,\,ds^2_\Y =  {{^{(\Y)}\!g}}_{\kk\ll} dy^\kk dy^\ll.
\ee

We have denoted  points of $\Y$ by $y$,
points of $\X$ by $x$  and those of $\M$ by $X=(x,y)$ (we
will use the same symbols for the corresponding coordinates);
 $\a,\b$ are tensor indices on $\X$, $\kk,\ll$ on $\Y$
and $\mu,\nu$ on $\M$. Notice that the Riemannian metric
$ {^{(\X)}\!g}$ is chosen with signature $(-,-,\cdots,-)$.

The pseudo-Riemannian (Lorentzian)  manifold $(\M,g)$  is  called a
{\em warped
product} \cite{O'Neill}; this structure is also denoted concisely
by writing
$
\M=\X\times_\omega \Y $.
$\M$ is therefore a (trivial) fiber bundle over
$\X$, whose fibers are all conformally equivalent to a manifold
$(\Y,{^{(\Y)}\!g})$ with a  conformal factor which depends only upon
the basis point $x$ \cite{tondeur}.\par

The simplest example of a warped product is provided by a
Minkowskian background geometry (in arbitrary dimension)
and it can be shown that its
warped product structure
can be realized in several ways, but with only two types of branes,
namely either
lower dimensional Minkowskian spacetimes or de Sitter spacetimes
(i.e. in geometrical terms: hyperbol\ae\  or one-sheeted hyperboloids).
This can be proven by a study of the Riemann tensor (see
\cite{BerGou1, BerGou2, BerGou3} for the relevant formul\ae\  in the
Riemannian case, which carry over to the pseudo-Riemannian as well
with obvious modifications). Similar remarks hold also for other constant
curvature spacetimes.

The Laplace--Beltrami operator for $0$-forms (functions)  on such a
manifold has the following structure:
\be
\square =\frac 1 {\sqrt{|g|}}\pa_\mu\left(\sqrt{|g|}
g^{\mu\nu}\pa_\nu\right)= \triangle_\X+{d} \left(\pa_\a \log(\omega)\right)\,
{^{(\X)}\!g}^{\a\b}\pa_\b+
\frac 1 {\omega^2} \square_\Y .
\ee
We will assume that $\M$ is globally hyperbolic and consider a {\em
canonical } quantum field $\PHI$ on $\M$ satisfying
 the Klein--Gordon equation
\be
(\square+M^2)\PHI(X)=\le(\tilde\triangle_\X +\frac
1{\o^2(x)}\square_\Y + M^2 \ri)\PHI(x,y)=0,
\ee
where we have  introduced  the operator
\be
\tilde\triangle_\X=\triangle_\X+
d \left(\pa_\a \log\omega\right)  {^{(\X)}\!g}^{\a\b}\pa_\b
=  \frac 1 {\omega^d\sqrt{|  {^{(\X)}\!g}|}} \pa_\a
\left(\sqrt{|  {^{(\X)}\!g} |} \,\omega^{d} \, \, { {^{(\X)}\!g}}^{\a\b}\pa_\b\right).
\ee
Separation of variables leads to  the following equations for the modes
\ba
&& \left(\square_\Y + \lambda\right)\varphi(y) = 0
\\
&&
\o^2(x) \left( \tilde\triangle_\X+ M^2     \right) \theta(x) =  \lambda   \theta(x).\label{eig}
\ea
Eq. (\ref{eig}) can be considered to define  a spectral problem
in the Hilbert space
\be
{\cal H} = {\rm L}^2(\X, d\tilde v_\X) ,\qquad d\tilde v_\X(x) =\omega^{d-2}(x) dv_\X(x),
\label{hilbert}
\ee
where $ dv_{_\X}(x)= \sqrt{|  {^{(\X)}\!g} |}  \ dx$ is the invariant volume
form on $\X$. Indeed, the  operator $\omega^2 (x) (\tilde\triangle_\X + M^2)$
 is symmetric on the dense domain  $C^\infty_0(\X)\subset {\cal H}$.
If we assume that  such operator has a self--adjoint extension (which
may or may not be the case in specific examples),
 the spectral theorem provides us with  a basis
$\{\theta^{(i)}_\lambda\}$  of  generalized
eigenvectors  which gives a decomposition of the identity.
In the same fashion as in the introductory example we then have
\bea
& &
(\theta_{\lambda}^ {(i)},\theta_{\lambda'}^{(j)}) = \int_{\X}
\overline\theta_{\lambda}^{( i)}(x)
\theta_{\lambda'}^{( j)}(x) {\rm d}\tilde v_\X =
\delta(\lambda-\lambda')\delta_{ij}\cr
& & \sum_{\lambda,i} \overline\theta_{\lambda} ^{(i)}(x)\theta_{\lambda}^{ (i)}(x') =
\omega^{2-d}(x)\delta_\X(x,x') , \label{comp}
\eea
where the indices $(i),(j)$ label the possible degeneracy of the
(possibly continuous) spectrum,
$\delta_\X(x,x')$ is the delta distribution on $\X$ and
the prefactor $\omega^{2-d}(x)$ comes from the definition of the
Hilbert product. \par
As in the toy-example treated in the previous section,
we introduce ``formal quantum fields'' $\Varphi_{\lambda}^{( i)}(y)$ by a
smearing out of the above modes:
\be
\Varphi_{\lambda }^{(i)}(y)=\int_{\X}
\PHI(X) \overline\theta_{\lambda}^{(i)}(x) {\rm d}\tilde v_\X(x) .
\label{hu}
\ee
Two remarks are in order here.\\
First, it is not obvious a priori that this expression makes
sense at all, since we are smearing an operator valued
distribution with a  function which does not belong to the
corresponding test function space. At best, the fields
$\Varphi_{\lambda }^{(i)}(y)$ can be operator-valued
distributions w.r.t.  $\lambda$ and $y$, namely, to get a {\em
bona fide} operator one should smear $ \Varphi_{\lambda
}^{(i)}(y)$ against suitable test functions in $\lambda$ and $y$
(as in the toy-example).\\
Second, while the Hilbert space $\cal H$ may seem to be
the most natural where to study
the eigenvalue problem given in Eq. (\ref{eig}), its  choice
is by no means mandatory. Different choices may  produce
different formul\ae.

 By formally  using  Eq. (\ref{comp})
we can invert the transformation (\ref{hu}) and get
\be
\PHI(X)=\sum_{\lambda,i} \theta_{\lambda }^{(i)}(x)
\Varphi_{\lambda}^{( i)}(y).
\ee
In concrete applications the actual viability of this inversion
needs to be verified case by case.\par
In the following we  use {\em real}--valued
eigenfunctions $\theta_{\l}^{(i)}$
so that the fields $ \Varphi_{\lambda }^{(i)}(y)$ are Hermitean.\\
Under the assumptions of self-adjointness that
we have postulated, the following properties hold: \\
{\em a)  The  fields
$\Varphi_{\lambda}^{( i)}$ satisfy the
 Klein--Gordon equation on the
manifold $\Y$ (in cases of interest to us, $\Y$ is Lorentzian).\\
b) The fields  $\Varphi_{\lambda}^{( i)}$ commute for
$\lambda\neq\lambda'$ or $i\neq j$.}\\
The proof of assertion {\em a)} comes from the following chain of equalities (in the sense
of distributions in $\lambda $ and $y$):
\bea
& & (\square_\Y+\lambda)\Varphi_{\lambda}^{( i)}(y) =\int_{\X}
{\theta_{\lambda}^{( i)}} (x)
(\square_\Y+\lambda)
 \PHI(X){\rm d}\tilde v_\X(x)=\cr
& & =\int_{\X}
{\theta_{\lambda}^{( i)}} (x)
[-\omega^2(x)(\tilde\triangle_\X+M^2)+\lambda]
 \PHI(X){\rm d}\tilde v_\X(x)=\cr
& &=\int_{\X} \left\{
[-\omega^2(x)(\tilde\triangle_\X+M^2)+\lambda]
\theta_{\lambda}^{( i)} (x)\right\}
 \PHI(X){\rm d}\tilde v_\X(x)=0,
\eea
where we made use of the assumed self--adjointness of the operator
$\o^2(\tilde\triangle_\X+M^2)$ (but not of the hermiticity of the fields).\\
The two-point correlation functions of the fields
$\Varphi_{\lambda}^{( i)}$ on the vacuum
$ {\Omega}$ of the field $\PHI$ is then   given by:
\bea
& & W_{\lambda,\lambda'}^{(ij)}(y,y') =\le\langle\Omega,
\Varphi_{\lambda}^{ (i)}(y)\Varphi_{\lambda'}^{( j)}(y')\Omega\ri\rangle\cr
& &=\int_{\X \times \X}\hspace{-15pt}
{\rm d}\tilde v_\X(x){\rm d}\tilde v_\X(x') \theta_\l^{(i)}(x)
\theta_{\l'}^{(j)}(x')W\le(X,X'\ri).
\eea
If we invert this formula by making use of (\ref{comp}),
we obtain the following representation
for $W$:
\be
W(X,X')=\sum_{\l,\l',i,j}\theta^{(i)}_\l(x)\theta^{(j)}_{\l'}
(x')W^{(ij)}_{\l,\l'}(y,y')\label{integr}
\ee
The distribution $ W_{\lambda\lambda'}^{(ij)}(y,y')$
satisfies the Klein--Gordon equation on $\Y$ w.r.t. both $y$ and $y'$,
with
masses  $\sqrt{\lambda} $ and respectively $\sqrt{\lambda'}$ .\par

\vskip 0.2cm

We now prove assertion {\em b)},  namely that the  quantum fields
$\Varphi^{(i)}_\lambda$ commute for different values of $\lambda$
or $i$. Indeed, the CCR's  for the field $\PHI$ can be written as
follows: \bea & & \le[\PHI(X),\PHI(X')\ri]_{\lfloor_{\cal
C}}=0 \label{CCRo} \\
& &\le[\PHI(X),{\pa_{t'}} \PHI (X')\ri]_{\lfloor_{\cal C}}
= i \delta_{\cal C}(X,X')\ ,
\eea
where $\pa_t$ denotes  a  time--like vector, orthogonal to  a
given Cauchy surface   $\cal
C$ and normalized to unity (this is not necessarily the gradient of a
time parameter).
We have adopted the following  convention:
 whenever we have a (Riemannian) submanifold $S
\hookrightarrow\M$, then $\delta_S(p,p')$ denotes the delta
distribution on that submanifold w.r.t. the volume element inherited
from the ambient manifold $\M$.\par
Taking advantage of the  structure of $\M$, we can choose
a Cauchy surface in
the form ${\cal C}=\X\times\Sigma$, where $\Sigma$ is a Cauchy
surface in  $\Y$; the former equations now read
\bea
& & \le[\PHI(X),\PHI(X')\ri]_{\lfloor_{\cal
C}}=0,  \label{CCRo1}, \\
& &\le[\PHI(X),{\pa_{t'}} \PHI(X')\ri]_{\lfloor_{\cal C}}
=i\delta_{\cal C}(X,X')
=i\omega^{1-d}(x)\delta_\X(x,x')
\delta_\Sigma(y,y')  ;\label{CCR}
\eea
the factor $\omega^{1-d}$
comes from the volume element of $\cal C$ which is given by  $dv_{\cal C} = \omega^{d-1}dv_\X\otimes
dv_\Sigma$ (recall that the surface $\Sigma$ has dimension $d-1$).
The vector $\pa_t$ is a time--like vector orthogonal to
${\cal C} =\X\times \Sigma$ and normalized w.r.t. the metric of $\M$:
it follows that the vector $\omega(x) \pa_t$ is a time--like vector
orthogonal to $\Sigma$ and normalized w.r.t. the metric in
$\Y$.\footnote{Indeed, let
$\pa_t$ be a normalized vector  tangent to $\M= \X\times_\o\Y$ at the
point $(x,y)$: then
its projection onto $\Y$ has norm $\o^{-2}(x)$, for
 $1=g(\pa_t,\pa_t)=\o^2(x){^{(\Y)}\!g}(\pa_t,\pa_t)$.}
We will denote by $\pa_T$ the vector $\omega(x)\pa_t$  tangent to  $\Y$ and also (with a slight
abuse of notation) its lift to the
tangent bundle of $\M$.
With this rescaling  Eq. (\ref{CCR}) reads
\be
\le[\PHI(X),{\pa_{T'}} \PHI(X')\ri]_{\lfloor_{\cal C}}
=i\omega(x)\delta_{\cal C}(X,X')
=i\omega^{2-d}(x)\delta_\X(x,x')
\delta_\Sigma(y,y') \label{CCRRES} .
\ee
We now smear both sides of Eq. (\ref{CCRo1}) and Eq. (\ref{CCRRES})
 with the modes $\theta_\l^{(i)}(x)$ and
$\theta_{\l'}^{(j)}(x')$ and apply Eq. (\ref{hu}):
Eq. (\ref{CCRo1}) gives an analogous equation for the fields
$\Varphi_l^{(i)}$ and Eq. (\ref{CCRRES}) gives
\bea
&& \le[\Varphi_\l^{(i)}(y),{\pa_{T'}} \Varphi_{\l'}^{(j)}
(y')\ri]_{\lfloor_{\Sigma}}=
\int_\X {\rm d}\tilde v_\X(x)  \int_\X {\rm d}\tilde v_\X(x')\theta_\l^{(i)}(x) \theta_{\l'}^{(j)}(x')
\omega(x)^{2-d} \delta_\X(x,x')\delta_\Sigma(y,y') =\cr
&&=
\delta_{ij}\delta(\l-\l') \delta_\Sigma(y,y')\ .\label{normal}
\eea
It follows that $\Varphi_{\lambda}^{( i)}$ commutes everywhere on
$\Y$ with $\Varphi_{\lambda' }^{(j)}$ for $\lambda\neq\lambda'$ or
$\lambda =\lambda'$ but $i\neq j$: in fact the above equations
tell that on the Cauchy surface $\Sigma$, for
$\lambda\neq\lambda'$, the Klein--Gordon fields
$\Varphi_\lambda,\ \Varphi_{\lambda'}$ commute between themselves
along with their canonical momenta and hence they do commute
everywhere in $\Y$ as a consequence of the equations of motion.
This ends the proof of assertion {\em b)}.\par

\vskip 0.3cm
In all the examples that we shall present (as in the toy-example of the previous section),
this commutativity of the formal fields will follow from a stronger property, namely
the diagonal character of their correlation functions
$ W_{\lambda,\lambda'}^{(ij)}(y,y'),$
which will be
of the form $\delta_{ij}\delta(\lambda-\lambda') W_{\lambda}(y,y')$.
This stronger property  may fail to be true in the generic case, unless
some additional structural properties on $\M$ are introduced.
This is precisely what will be done in our next section, in such a way
that all our examples are covered .

Whenever the previous diagonal form of
$ W_{\lambda,\lambda'}^{(ij)}(y,y')$
is valid, Eq.(\ref{integr}) immediately yields the corresponding diagonal decomposition:
\be
\bra{\Omega}\PHI(X)\PHI(X')\ket{\Omega} = \sum_{\lambda,i}
\theta_\l^{(i)}(x)\theta_\l^{(i)}(x')
W_{\lambda}(y,y')
\label{symbolic}
\ee
Moreover,
when we consider the field $\PHI$ restricted to a fixed slice $x=const$,
we obtain a superposition of Klein--Gordon fields
as an immediate consequence of the previous formula, namely:
 \be
\bra{\Omega}\PHI(x,y)\PHI(x,y')\ket{\Omega} = \sum_{\lambda,i}
|\theta_\l^{(i)}(x)|^2
\label{1}
W_{\lambda}(y,y')
\ee
This formula is analogous to the K\"allen--Lehmann representation for the
two--point function in the Minkowskian spacetime \cite{kallen}.\\
>From (\ref{1}) it follows that the weight function of this
K\"allen--Lehmann decomposition of the restricted
propagator is:
\be
\mu^{(i)}(\lambda,x)=\sum_{\lambda',j}
\delta_{ij} \delta(\lambda-\lambda')|\theta^{(j)}_{\lambda'}(x)|^2
\ee
which is the discontinuity of the resolvent of the
operator $\omega^2(x)(\tilde\Delta_\X + M^2)$ on its spectrum, i.e. the
density of states per unit spectrum per unit volume (in
$\X$).\vskip 5pt
If  $\X$ is a one--dimensional spatial
manifold we may take one step further.\\
Let us choose a  coordinate
$x$ such that  the line element on $\X$
is simply $-dx^2$.
The  spectral problem now leads to
\bea
& &\omega^2(x)\left(\varphi''(x)+d\frac{\omega'(x)}{\omega(x)}\varphi'(x)
-M^2\varphi(x)\right)= -\lambda \varphi(x)
\eea
where the Hilbert space has the inner product
\bea
& &(\varphi,\psi) =\int_{\X} {\rm d}x\, \omega^{d-2}(x) \overline\varphi(x)\psi(x)\ .
\eea
The transformation
\be
\varphi(x)=\omega^{\frac{1-d}2}(x) f(x)\label {lommel}
\ee
 allows us to rewrite the eigenvalue equation and the Hilbert product
as follows:
\bea
& &  f''(x)+\frac{\omega'(x)}{\omega(x)}f'(x)+\left[\frac{\lambda}{\omega^2(x)}
-M^2+\frac{1-d}2 \frac{\omega''(x)}{\omega(x)} -\frac{(d-1)^2}4 \left( \frac
{\omega'(x)}{\omega(x)} \right)^2   \right] f(x)=0\cr
& &(f,h)=\int_{\X}\frac {{\rm d}x}{\omega(x)} \overline f(x)h(x)\ .
\eea
Let us introduce a coordinate $s$ so that
\be
ds=\frac{dx}{\omega(x)}\ .\label{change}
\ee
We obtain that:
\bea
& &-f''(s)+U(s)f(s)=\lambda f(s)\cr
& &(f,h)=\int_{\X} \overline f (s) h(s)\, {\rm d}s \ ,
\eea
where
\bea
U(s) =  \frac {d-1}2
\frac{\o''(s)} {\o(s)}+\le(\frac{\o'(s)}{\o(s)}\ri)^2\le[
\frac{(d-1)^2}4+\frac {1-d}2 \o(s)\ri] +M^2\o^2(s)\ ,
\eea
and prime now means derivative w.r.t. the variable $s$. \par
We have obtained a one dimensional Schr\"odinger problem with a
potential $U(s)$ which depends on the warping function $\o(s)$.  Notice
that the result matches the introductory example  for the flat case;
this is a trivial instance of the above general framework, where
$\X=\R$, $\Y=\R^d$ and $\omega(x)=1$:  the operator
$\omega^2(x)(\tilde\triangle_\X+M^2)=-\pa^2_x+M^2$ describes exactly a
free Schr\"odinger particle with constant potential
$M^2$.
\section{Warped manifolds with additional geometrical structure.}
In order to give relevant applications of the previous theoretical
setting, we need to specify additional structural properties on
the geometry of the warped manifold $\M$. These geometrical
properties will be sufficient to establish (via the lemma stated
below) the {\em validity of the diagonal decomposition}
(\ref{symbolic}) which then entails the existence of a
K\"allen-Lehmann-type decomposition for the bulk Klein-Gordon
fields built in terms of a ``tower'' of massive fields living on
the brane $\Y$.

Such geometrical properties involve appropriate {\em consistency
requirements} between the geometry of $\M$ and that of the leaves
$\Y_x$ that deal with global symmetries as well as with the
existence of complexified manifolds for $\M$ and $\Y$.

\vskip 0.4cm \noindent {\em i) Consistency of  global isometries of the leaves.}

We assume that there exists an isometry group $G$ of $\M$   and a
subgroup\footnote{It is not necessary that $G$ is a {global} isometry
group of $\M$; $G$ can be identical to $G_{\Y}$, as it will occur in
most of the applications below.  However, in the latter there will be a
larger global isometry group acting on an {\em extension} $\hat \M$ of
$\M$ on which the ambient two-point function $W(X,X')$ is defined and
admits this global isometry.} $G_{\cal Y}$ of $G$ acting in each leaf
${\cal Y}_x$ of ${\cal M}$ as a global isometry group of ${\cal Y}$ and
that there exists a {\rm global} pseudo-distance $z(y,y')$ on $\Y$
which is preserved by this isometry group $G_{\Y}$.

\vskip 0.4cm \noindent {\em ii) Consistency of complex
geometries.}

We assume that $\M$ and $\Y$ admit respective complexified
manifolds ${\cal M}^{(c)}$ and ${\cal Y}^{(c)}$, such that for
each $x$ in $\X$ the complexified ${\cal Y}^{(c)}_x$ of $\Y_x$ is
contained in ${\cal M}^{(c)}$. Moreover  ${\cal M}^{(c)}$ and
${\cal Y}^{(c)}$ contain distinguished pairs of domains, called
respectively the {\em tuboids} $T^{\pm}$ and ${\cal T}^{\pm}$ in
such a way that for all $x$ in $\X$, one has:
\begin{equation}
{\cal T}^+_x \subset T^+\ \ {\rm  and}\ \ \  {\cal T}^-_x \subset T^-.
\label{inclusion}
\end{equation}
These tuboids $T^{\pm}$ (resp. ${\cal T}^{\pm}$) serve to define
a preferred class of (generalized) free fields on $\M$ (resp.
$\Y$), as being those whose two-point functions are boundary
values of holomorphic functions ${\rm W}(X,X')$ (resp. ${\rm
W}(y,y')$) in the product domain $T^- \times T^+$ (resp. ${\cal
T}^- \times {\cal T}^+ $). This property, which is a
generalization of the standard analyticity property of Minkowskian
two-point functions in the Wightman axiomatic framework, is called
{\em normal analyticity} (see its introduction in the de Sitter
case in \cite{Bros:1996js} and more recently its extension to the
anti-de Sitter case in \cite{BBMS:1999}).

On the basis of the previous consistency requirements, we shall
now establish the following statement (where we have kept the
notations of the previous section, but dropped the discrete
variables $i,j$): \\
{\bf Lemma:} {\sl Consider the distribution in $(\lambda,\lambda')$
defined as the two-point function of the  
formal fields $\hat \varphi_{\lambda}(y)$
and $\hat \varphi_{\lambda'}(y')$, namely 
$$ W_{\lambda,\lambda'}(y,y') =\le\langle\Omega,  
\Varphi_{\lambda}(y)\ \Varphi_{\lambda'}(y')\Omega\ri\rangle $$ 
\begin{equation}
=\int_{\X \times \X}\hspace{-15pt} 
{\rm d}\tilde v_\X(x)\ {\rm d}\tilde v_\X(x') \theta_\l(x) 
\theta_{\l'}(x')\ W\le(X,X'\ri).
\label{formaltwopoint}
\end{equation}
where $W(X,X')$ denotes the two-point function of a Klein-Gordon field $\PHI (X)$ 
on $\M$ satisfying $G-$invariance and normal analyticity in ${\cal
M}^{(c)}$, and the integral over  
$x$ and $x'$ in (\ref{formaltwopoint}) is supposed to be convergent after
smearing out in the variables $\lambda, \lambda'$ for all  
real or complex values of $y$ and $y'$ (in ${\cal T}^- \times {\cal T}^+$). 
\vskip 0.2cm

Then the distribution 
$ W_{\lambda,\lambda'}(y,y')$    
is of the following diagonal form
\begin{equation}
W_{\lambda,\lambda'}(y,y')=   \delta(\lambda-\lambda') 
W_{\lambda}(y,y'),   
\label{diag}
\end{equation}
where 
$W_{\lambda}(y,y') = w_{\lambda}(z(y,y'))$ is a solution of the 
Klein-Gordon equation (in both variables $y, y'$)  
$$ \square_y 
W_{\lambda}(y,y') = 
\square_{y'} 
W_{\lambda}(y,y') = -\lambda  
W_{\lambda}(y,y') $$ 
satisfying the required properties of a two-point function on $\Y$, namely  
$G_{\Y}-$invariance and normal analyticity property
in ${\cal Y}^{(c)}$. Moreover  
$W_{\lambda}(y,y')$ is correctly normalized, in consistency 
with the canonical commutation relation for the corresponding Klein-Gordon field, 
namely one has (with the notations of section 2):
$${\pa_{T'}} 
\left(W_{\lambda}(y,y') - W_{\lambda}(y',y)\right) 
_{\lfloor_{\Sigma}}=  
\delta_{\Sigma}(y,y').$$}

To show this lemma we observe that, in view  of (\ref{formaltwopoint}), the invariance of $W(X,X')$ under $G$ implies the 
invariance of 
$ W_{\lambda,\lambda'}(y,y') $ 
under $G_{\cal Y}$ . 
The latter is therefore of the form 
$ W_{\lambda,\lambda'}(y,y') 
=  w_{\lambda,\lambda'}(z(y,y'))$  
and in view of the symmetry of the distance  
w.r.t. $y$ and $y'$, one has {\sl in the sense of distributions in $(\lambda,\lambda')$}:
$$\square_y W_{\lambda,\lambda'}(y,y') 
- \square_{y'}  W_{\lambda,\lambda'}(y,y') =0$$
and therefore, {\em in view of property a) of the fields} $\hat \varphi_{\lambda}(y)$:  
$$(\lambda - \lambda') W_{\lambda,\lambda'}(y,y') =0,$$
which entails that
$W_{\lambda,\lambda'}(y,y')$ is of the form (\ref{diag})  
(since the general solution as a distribution of the equation $x_1 T(x_1,x_2) =0$ is 
$T(x_1,x_2)= \delta(x_1)\times t(x_2)$).
The normal analyticity of  
$W_{\lambda}(y,y')$ results from the normal analyticity of $W(X,X')$ in view of the    
inclusion relations (\ref{inclusion}) and the assumed convergence of the integral in 
(\ref{formaltwopoint}).   
Finally, the normalization of 
$W_{\lambda}(y,y')$     
readily follows from the  
commutation relation (\ref{normal}) established in section 2   
by integrating the latter over $\lambda'$.

\section{Applications.}
 In the four examples studied below, we  discuss quantum
field theories on manifolds which admit natural complexified
manifolds carrying tuboids of normal analyticity, and in all
these theories the geometric symmetries are unbroken, namely the
considered two--point functions $W(X,X')$ are invariant under the
global isometries of the ambient manifold $\M$; moreover, the
leaves $\Y_x$ will always satisfy the two geometrical consistency
requirements specified above.

In the first two examples, $\Y$ is a de Sitter spacetime and
$\X$ will be the half line or the segment $(0,\pi)$ with
appropriate measures; they give  a structure of warped product to open
subsets of the Minkowski space in the first example, and to an ambient
de Sitter spacetime in the second example: this extends and
generalizes the results in \cite{Bertola:1999ga}.

In the third example we will
revisit the Unruh problem, namely the restriction of an ambient
Minkowskian quantum field theory to the world-line $\Y$ of a uniformly accelerated
observer. In this case,
the isometry group of $\Y$ (induced by a Lorentz boost subgroup of the ambient space)
is simply the time translation group in the proper time of the accelerated
observer.

The last example regards QFT on the anti--de Sitter manifold,
considered as foliated by Minkowskian branes. Although this case
seems to lie out of the picture drawn in section 2,  because AdS
spacetime is not globally hyperbolic, it turns out that this lack
of global hyperbolicity is not an obstruction to the applicability
of the previous lemma. In fact, the previous geometrical
consistency requirements are still fulfilled there.

In all these examples, the diagonal form (\ref{diag}) of the
correlators $W_{\lambda,\lambda'}^{(ij)}(y,y')$ of the formal
fields $\Varphi_{\lambda}^{( i)}$, $\Varphi_{\lambda'}^{( j)}$
will always allow us to interpret each formal field
$\Varphi_{\lambda}^{( i)}$ as a genuine Klein--Gordon field with
the corresponding two-point function $ W_{\lambda}(y,y')$ on the
brane, and to obtain thereby, via the inversion argument given in
Section 2 (based on the completeness relation (\ref{comp})), a
decomposition of the ambient Wightman function $W(X,X')$ with the
diagonal form (\ref{symbolic}).

The consistency requirements which we consider in this section
readily imply (without any computation) that the restriction to
any given leaf $\Y_x$ of any Klein--Gordon field of the ambient
space $\M$ is a generalized free field on this leaf. When the
branes are either Minkowski or de Sitter spacetimes, as in the
examples we will present, there exists also a direct method for
computing the spectral function of this restricted field by a
Laplace-type transformation on the leaf (this is standard for the
Wightman fields in Minkowski space \cite{Streater} and has been
carried out for de Sitter fields in \cite{Bros:1996js} by using
the results on ``invariant perikernels on the one-sheeted
hyperboloids'' of \cite{BV}). It is to be expected that the
comparison of such Laplace-type expressions of the spectral
function with the one obtained here by the (completely different
and more general) warped-manifold method in terms of the
``Schr\"odinger modes'' $\theta_{\lambda}^i$ will provide new
interesting identities relating Hankel-type and Legendre-type
functions.

 Concerning the more technical problem of the convergence of the
(a priori formal) integrals and sums (\ref{formaltwopoint}) and
(\ref{symbolic}), we shall check the latter in all the examples
and prove in particular that (\ref{symbolic}) can be given a
well-defined meaning as an integral w.r.t. a suitable measure
over the allowed mass spectrum and possibly a sum over the
degeneracy indices. To this end we shall analyze the spectral
problem along the
general lines drawn in section 2.\\
In the first three examples the operator
$T_\X =\o^2(x)\le(\widetilde\triangle_\X+M^2\ri)$ {\em is} essentially
self--adjoint in  $L^2(\X,d\tilde v_\X)$  on the domain
$C^\infty_0(\X)$ because it reduces to ordinary Schr\"odinger
operators with smooth potentials  bounded from below; therefore we will not
discuss their self--adjointness, since this follows from general theorems
(see e.g. \cite{Reed}).\\
On the contrary, in the last anti--de Sitter
example the relevant operator is
essentially self--adjoint only for values of $M^2$ bigger than a certain
threshold ${M_0}^2$; below ${M_0}^2$ the operator is {\em not}
essentially self--adjoint but can be extended to a self adjoint
operator in many different ways. Among the infinite a priori allowable
extensions, two of them are of special relevance to the so--called AdS--CFT
correspondence \cite{BBMS:1999}.

\subsection{Decomposition of (bulk) Minkowski fields into de Sitter (brane) fields.}
 \label{app1}
In this example the manifold $\M$ is the set of all points 
which are space--like w.r.t. a given event, chosen as the 
origin in a $(d + 1)$-dimensional 
Minkowski spacetime, endowed with a system of inertial
coordinates  denoted by $\{X^\mu\}$, 
$\mu=0,\dots, d$.\\
The region $\M=\{X:\ X^\mu X_\mu<0 \}$ 
is foliated by a family of   
$d$-dimensional de Sitter spacetimes, identified with 
the hyperboloids   
$$ X\cdot X \equiv \eta_{\mu\nu} X^\mu X^\nu  = (X^0)^2-(\vec X)^2 = -R^2\ .$$ 
$\M$ has the structure of a  warped manifold with 
base  $\X=\R^+$ with coordinate 
$R$; the  fiber $\Y$ can be identified with
 a $d$-dimensional de Sitter spacetime 
with radius $R=1$; using a polar--like parametrization for the events
of $\M$, $X=R\, y$ with  $y^2=-1$,
the Minkowskian metric of $\M$ can then be rewritten as follows:
$$ 
ds^2=-dR^2+R^2 ds_{\Y}^2\ , 
$$ 
where $ds_{\Y}^2$ is the de Sitter  metric of $\Y$,  
obtained as restriction of the Minkowski metric of the ambient space.
This realizes $\M$ as a warped product with
warping function $\omega(R)=R$. \par
The operator  $\tilde\Delta_\X$ equals  
$-\pa_R^2-\frac d R\pa_R$  and  
we are led to  the following 
eigenvalue equation for the modes $\theta_\l$:
\be  
R^2\left(-\pa_R^2-\frac d R \pa_R+M^2\right)\theta_\l (R)=\lambda 
\theta_\l (R).\label{desittertransverse} 
\ee
The operator at the L.H.S. 
is essentially self-adjoint on the dense domain ${\mathcal C}^\infty_0$     
of the Hilbert  space  $L^2(\X,d\tilde{v}_\X)$, whose scalar product 
has the following explicit form:
\be
(\varphi,\psi)=\int_{\X}\overline\varphi (R)\psi 
(R)R^{d-2} {\rm d}R.
\ee 
By means of the  transformation (\ref{lommel}) and by rescaling 
$\rho=MR$ (which together are particular instances of the so called 
`` Lommel's transformations''), 
Eq. (\ref{desittertransverse}) 
is  turned  into the modified Bessel's equation.
By further introducing the new variable $M\,R=e^s$ we finally 
  obtain: 
\be 
-f_\l''+(e^{2s}-\nu^2)f_\l=0,
\ee \label{expowall} 
{with}
\be
 \nu^2 = \lambda-\frac {(d-1)^2}4. \label{expowall2} 
\ee 
The prime now means derivatives w.r.t. the variable $s$. 
This operator is now self-adjoint w.r.t. the standard $L^2$ product $\int_{\R} \overline f(s) 
h(s)\, {\rm d}s$. 

We have thus obtained  a  Schr\"odinger problem for a 
potential $e^{2s}$. The corresponding spectrum is absolutely 
continuous and
nondegenerate; it coincides with 
the positive real line. 
This implies the condition  $\lambda>(d-1)^2/4$. 

The solutions which have the correct asymptotic behavior at $s=\infty$
are $K_{i\nu}(e^s)$, where $K_{i\nu}(z)=K_{-i\nu}(z)$ denotes the
modified Bessel function \cite{Bateman}; it is   real  for real $\nu$.
The normalization can be obtained by studying the asymptotic behavior
at $s=-\infty$, where these solutions behave as free waves.

The final  result, expressed in the original coordinate $R$, is  the
following family of normalized generalized eigenfunctions:
\be
\theta_\lambda(R)=N_\lambda R^{\frac {1-d}2}K_{i\sqrt{\lambda-(d-1)^2/4}}(MR)\ ,\,\,\,\,\,\,
N_\lambda \equiv \frac 1 \pi \sqrt{\sinh\le(\pi \sqrt{\lambda- {(d-1)^2}/4}\ri)}\ .
\label{rmodes}
\ee
There hold the completeness and orthonormality relations:
\bea
\int_{\frac{(d-1)^2}{4} } ^\infty {\rm d}\lambda\,
\theta_\lambda(R)\theta_\lambda(R') = R^{-(d-2)}\delta(R-R'),\cr
\int_{\R^+} {\rm d}R\, R^{d-2} \theta_\lambda(R)\theta_{\lambda'}(R) =
\delta(\lambda-\lambda')\ .
\eea 
We now introduce the fields $\Varphi_{\lambda}(y)$ on the de Sitter  
manifold  $\Y$ by smearing the field $\PHI$ against the radial modes
(\ref{rmodes}), as in Eq.(\ref{hu}).
The main result  of this section is the following: \\[5pt]
{\em 
c) the fields $\Varphi_{\lambda } (y)$ 
correspond to de Sitter Klein--Gordon fields in the ``Euclidean'' \cite{Gibbons:1977mu}
(also called Bunch-Davies \cite{bunchdavies}) vacuum state, namely
the vacuum expectation value (v.e.v.) 
of $\Varphi_{\lambda } (y)$  on $\Y$ is given by 
\be
W_{\lambda,\lambda'}(y,y')= 
\langle \Omega|\Varphi_\l(y)\Varphi_{\l'}(y')|\Omega\rangle = \delta(\l-\l') W_\l^{(E,d)}(y,y'),
\label{restrizione}
\ee
where $W_\l^{(E,d)}$ is the $d$-dimensional Euclidean (Bunch--Davies)  
two-point function, equipped with its normal analytic structure  \cite{Bros:1996js}.
\label{thmds}
Moreover each Klein--Gordon field of the ambient Minkowski space (with arbitrary positive mass $M$)
admits the following expansion of its two-point function: 
\be
\langle \Omega, \PHI(X)\PHI(X') \Omega\rangle=
\int_{\frac{(d-1)^2}4} ^\infty {\rm d}\lambda
\theta_\lambda(R)\theta_{\lambda}(R') W_\lambda^{(E,d)}(y,y')\ .\label{expans}
\ee
with $\theta_\lambda(R)$ given by formula (\ref{rmodes}).  

This equation allows us to consider the quantum field $\PHI$ {\em
restricted} to a fixed  de Sitter brane  $R=R'$; 
it has the structure of a K\"allen--Lehmann expansion expressing 
the ambient quantum field $\PHI$ as a superposition of de
Sitter quantum fields on the brane
$\Varphi_\lambda$ with mass spectrum
\be
\langle \Omega,\PHI(X)\PHI(X') \Omega\rangle_{|_{R=R'}} =
\int_{\frac{(d-1)^2}4} ^\infty {\rm d}\lambda
|\theta_\lambda(R)|^2 W_\lambda^{(E,d)}(y,y')\ .
\ee}   

\vskip 0.3cm
The proof of the previous statement goes as follows: according to \cite{Bros:1996js}, both geometrical consistency requirements   
defined above are satisfied by the subset $\M$ of Minkowski space and the de Sitter leaf $\Y$:
the isometry group $G = G_{\Y}$ is the corresponding Lorentz group $SO_0 (1,d)$ and 
the tubes ${\cal T}^{\pm}_R$ are the intersections of the complex quadric ${\Y}_R^{(c)}$ 
with the tubes $T^{\pm}$ of the complexified Minkowski space ${\M}^{(c)} =  {\bf C}^{d+1}$.  
It follows that the previous lemma is applicable, and that we only  
have to check that the two-point function $W_{\lambda}(y,y')$ of formula (\ref{diag}) 
coincides in the present case with the function $W_\l^{(E,d)}$. 
Let us recall that  $W_\l^{(E,d)}$
is  a distribution in the de Sitter invariant variable 
$v= y\cdot y'$ which satisfies the de Sitterian Klein-Gordon equation 
with eigenvalue $-\lambda$ in both variables $y,y'$ and is given by
\be
W_\lambda^{(E,d)}(y,y') = C_{d,\nu} 
{P}^{(d+1)}_{-\frac{d-1}2+i\nu}(y\cdot
y')\ , \label{wbm}
\ee 
where 
\bea
&&C_{d,\nu} = \frac{\Gamma\le(\frac {d-1}2+i\nu\ri)\Gamma\le(\frac
{d-1}2-i\nu\ri)} { 2^d\Gamma\le(\frac d 2\ri) \pi^{\frac d 2}},\cr
&& {P}^{(d+1)}_{-\frac{d-1}2+i\nu}(v) = 2^{\frac {d-2}2}
\Gamma\le(\frac  d 2 \ri) (v^2-1)^{\frac {2-d}4}
P^{\frac{2-d}2}_{-\frac 12 -i\nu}(v)\ ,\label{eucvac}
\eea
and $P^a_b$ denotes the associated Legendre function \cite{Bateman}.\par
The value of the constant $C_{d,\nu}$ ensures the correct normalization of   
$W_\l^{(E,d)}$, the canonical commutation relations being satisfied 
by the corresponding Klein--Gordon field. 
Moreover this distribution is correctly defined as being the 
{\em boundary value of the holomorphic function
${P}^{(d+1)}_{-\frac{d-1}2+i\nu}(y\cdot
y')$  from the tuboid $\{ (y,y') \in {\cal T}^- \times {\cal T}^+\}$ of  
${\Y}^{(c)} \times {\Y}^{(c)}$} \cite{Bros:1996js}.  

So by all its properties, this definition of $W_\l^{(E,d)}$ coincides with   
that of $W_{\lambda}(y,y')$ given in the lemma, which proves formula (\ref {restrizione}), 
and therefore the rest of property {\em c)} (in view of (\ref{comp})).  

It is worthwhile to remark that, when explicitly written
Eq. (\ref{expans}) is a rather complicated new integral relation between 
Legendre and Hankel functions. Here we get a ``quantum field theoretical''
proof of  that integral relation without actually performing any integral.

\vskip 1cm 
It is interesting to derive an alternative expression for  
$ W_{\lambda,\lambda'}(y,y') $  
by plugging  the momentum representation of the Minkowskian two-point function 
$W(X,X')$  into its defining formula (\ref{formaltwopoint}). We obtain: 
\bea
&& W_{\lambda,\lambda'}(y,y') = \cr
&& 
  \!\! \!\! \!\! \!\!\!\! \!\! \!\! \!\! = \int_0^\infty \frac {{\rm d}R}R R^{d-1} \theta_\lambda(R) \int_0^\infty
\frac {{\rm d}R'}{R'}  {R'}^{d-1} \theta_{\lambda'}(R')  \int \frac{{\rm
d}^{d+1}P}{(2\pi)^{d}}
\delta(P^2-M^2)\Theta(P_0) e^{-iP(X-X')}.
\label{expr}
\eea
In this expression we insert  the  parametrizations 
$X=R\,y$ and $X'=R'\,y'$  and introduce a vector $\alpha$ defined
by the relation 
$M \alpha = P$, so that $\alpha $ varies on the unit shell.  
One then rewrites the subintegral over $P$ as   
$$   \int \frac{{\rm
d}^{d+1}\alpha}{(2\pi)^{d}}
\delta(\alpha^2-1)\Theta(\alpha_0) e^{i(y'-y)\cdot \alpha M R}.$$ 
and by exchanging the order of the integrations over $R,R'$ and $\alpha$  
one is led to introduce the following integrals:
\bea
&& \varphi_\lambda(y,\alpha) = \varphi_\lambda(y\cdot \alpha) = 
M^{\frac {d-1}2} \int_0^\infty e^{i y\cdot\alpha\, MR} \theta_\lambda(R)R^{d-1}
\frac {{\rm d}R}R = \cr
&& = \sqrt{\frac \pi 2} 
 N_\lambda \Gamma\le(\frac {d-1}2-i\nu
\ri)\Gamma\le(\frac {d-1}2+i\nu\ri)\le((-iy\cdot\alpha)^2-1\ri)^{\frac{2-d}4}
P^{\frac{2-d}2}_{-\frac {1}2 -i\nu}(-iy\cdot\alpha).\label{plane} 
\eea
The functions $\varphi_\lambda(y,\alpha)$  
are plane waves  on de Sitter manifold i.e. are modes satisfying  the
de Sitter Klein--Gordon equation whose phase 
is constant on planes; $\alpha$  plays the role of wave-vector.
$P $ is an associated Legendre function \cite{Bateman}. 
The integral appearing in the definition of these waves 
is well defined at 
both extrema provided $|\Im(\nu)|< \frac {d-1}2$.\\
Then by rewriting the integral (\ref{expr}) in terms of these waves, we obtain the following  
new integral representation
for the Bunch-Davies de Sitter two-point function:
$$W_{\lambda,\lambda'}(y,y')= \delta(\lambda-\lambda') 
C_{d,\nu} 
{P}^{(d+1)}_{-\frac{d-1}2+i\nu}(y\cdot y') \ $$ 
\be
= \int \frac {{\rm d}^{d+1}\alpha}{(2\pi)^{d}} \delta(\alpha^2-1)\Theta(\alpha_0)
{\overline \varphi}_{\lambda} ( y\cdot \alpha ) \varphi_{\lambda'}(y'\cdot\alpha )\ .
\label{newexpr}
\ee
\subsection{Decomposition of (bulk) de Sitter fields into lower dimensional
(brane) de Sitter fields.}
\label{dS2dS}
In the second example we are dealing with a family of
$d$--dimensional de Sitter branes  embedded in a
$(d+1)$--dimensional de Sitter spacetime. As explained in
\cite{moschellaschaeffer} this problem is physically relevant to
understand the spectrum of the density fluctuations in an open
inflationary cosmology.

Let us consider a $(d+2)$--dimensional Minkowski spacetime, with a
chosen set of inertial coordinates $X^0,...,X^{d+1}$. The bulk de
Sitter manifold is taken to be the unit one-sheeted hyperboloid
$DS=\{X\in{\mathbb M}^{d+2},\ X\cdot X=-1\}$. Consider now the
following open region of the bulk: $\{X\in DS: \ |X^{d+1}|<1\}$. This
region is foliated by $d$-dimensional de Sitter branes,  obtained
by intersecting the bulk with a family of hyperplanes
parameterized by a coordinate $x\in (0,\pi)$ as follows:
$\{X\in{\mathbb M}^{d+2}, X^{d+1}=\cos x\}$.

The metric of the bulk de Sitter manifold can consequently be
written as follows: \be ds^2_{DS}=-dx^2+\sin^2 x \ ds^2_{dS}; \ee
$ds^2_{dS}$ is the metric of a $d$--dimensional de Sitter
manifold with radius $R=1$, and $\omega(x)=\sin x$. The base
manifold $\X$ is thus the segment $(0,\pi)$ with coordinate $x$
and metric $dx^2$.

The spectral problem now is the following: \be \omega^2(x)
(\tilde\triangle_\X+M^2)\theta_\l= \sin^2 x \ \theta_\l''+d
\cos^2 x \ \theta_\l'-\sin^2 x \ M^2 \theta_\l= -\lambda\theta_\l
; \ee it has to be considered in the Hilbert space whose product
is $(\varphi,\psi) = \int_0^\pi \le(\sin
 x \ri)^{d-2}\overline \varphi (x)
\psi (x)\, {\rm d}x$. $M$ is the mass of the field propagating in
the ambient de Sitter space.

Following Eq. (\ref{lommel}), this equation can be simplified by
introducing  $\theta(x)=\sin^{\frac{(1-d)}2} x  f(x)\ .$ A
further  simplification is achieved by introducing the coordinate
$s = \mbox{arc} \tanh
 \cos x$. The operator and the inner
product become \bea & &-f''(s)+
\frac{M^2-\frac{d^2-1}4}{\cosh^2(s)}
f(s)=\left(\lambda-\frac{(d-1)^2}4\right) f(s), \\
& &(f,h)=\int_\R {\rm d}s\, \overline f(s)h(s)\ , \eea where
again the prime denotes the derivative w.r.t. $s$. We have
obtained a Schr\"odinger problem with potential $U(s)=
\frac{M^2-\frac{d^2-1}4}{\cosh^2(s)}$: this is either a barrier
or a well according to the sign of $M^2-\frac {d^2-1}4$. When
this quantity is negative some bound states may appear depending
on the depth of the well. In both  cases a positive continuous
doubly--degenerate spectrum will persist, for which
$\lambda-\frac{(d-1)^2}4=q^2\geq 0$.\par It is now a standard
quantum mechanical problem to find eigenfunctions and eigenvalues
of this Sch\"rodinger problem.
\paragraph{Continuous spectrum:}
the continuous spectrum coincides with the positive real axis. We
will write $q^2= \lambda-\frac{(d-1)^2}4$ with $
\lambda>\frac{(d-1)^2}4$ for the eigenvalue. 
Using standard techniques
for the study of Schr\"odinger operators one can find 
the following family of orthonormal {\em
complex} generalized eigenfunctions labeled by a positive parameter
$q$:  
\bea 
&& F_q(x)=\frac{e^{\frac{\pi q}2}\Gamma(1+\rho-iq)\Gamma(-\rho-iq)}
{\sqrt{2\pi} \Gamma(-iq)}\le(\sin x\ri)^{\frac {1-d}2}
 P^{iq}_\rho(\cos x +i\epsilon),\cr
&& F_{-q}= \frac{e^{\frac{-\pi q}2}\Gamma(1+\rho-iq)\Gamma(-\rho-iq)}
{\sqrt{2\pi} \Gamma(-iq)}\le(\sin(x)\ri)^{\frac {1-d}2}
 P^{iq}_\rho(-\cos x -i\epsilon),
\eea where $\rho$ satisfies \be\rho(\rho+1)=\frac{d^2-1}4-M^2 =
-\frac 14 -\nu^2 \ee (see Eq. (\ref{eucvac})) so that $ \rho=
-\frac 1 2 + i\nu$. The required {\em real} modes
$\theta_{q,\epsilon}$ with $\epsilon \in \{ s,c \}$ are given by
\bea &&\theta_{q,s}(x)=\frac {1}{2i}\le(
F_q(x)-\overline{F_q(x)}\ri),\cr &&\theta_{q,c}(x)=\frac
{1}{2}\le( F_q(x)+\overline{F_q(x)}\ri). \eea
\paragraph{\bf Discrete Spectrum:}
When $M^2-\frac{d^2-1}4 <0 $,  bound states can exist. We can
construct them
 by the substitution $q\to
-iq$ in the formul\ae\ for the generalized eigenfunctions
corresponding to the continuous spectrum; now $\rho$ is solution
of $$(\rho+1)\rho=\frac{d^2-1}4 -M^2\ . $$ We fix the root
$\rho=-\frac 1 2 +\sqrt{\frac{d^2}4-M^2}>0$.
Standard quantum mechanics then says that the discrete eigenvalues
are \be q_n = -n-\frac 1 2+\sqrt{\frac {d^2}4-M^2}>0. \ee
Consequently, the number of bound states is \be \#\{\hbox{Discrete
Spectrum}\}=\left[-\frac 1 2+\sqrt{\frac {d^2}4-M^2}\right]\ , \ee
where the square brackets denote the integral part. The
normalization of the corresponding states can be computed using
the following integral \cite{gradshteyn} \be \int_{-1}^1
\frac{{\rm d}y}{1-y^2}\,\le|P^{-q}_\rho(y)\ri|^2=\frac
{\Gamma(1+\rho-q)}{q\Gamma(1+\rho+q)}\ . \ee
It follows that the normalized eigenfunctions are
\be
\theta_n(x)=\le(\sin(x)\ri)^{\frac {1-d}2}
P^{-\rho+n}_\rho(\cos(x)+i\epsilon)\sqrt{\frac{(\rho-n)
\Gamma(1+2\rho-n)} {n!}},
\ee
with $n=0,\dots,[\rho]=\left[
-\frac 1 2+\sqrt{\frac{d^2}4 -M^2}\right]$.
As before, let us introduce the formal quantum fields \beas &&
\Varphi_{q,\epsilon} (y) = \int_{0}^\pi {\rm
d}x\,\le(\sin(x)\ri)^{d-2} \theta_{q,\epsilon} (x) \PHI(x,y) \ ,\
\ \ q\in \R^+,\ \epsilon\in\{s,c\} \cr && \Varphi_n(y) =
\int_{0}^\pi {\rm d}x\,\le(\sin(x)\ri)^{d-2} \theta_n(x)
\PHI(x,y)\ . \eeas
By the same arguments used in Section (\ref{app1}) 
we obtain that:
\vskip10pt
{\em d) The fields $\Varphi_{q,\epsilon } (y)$, $\Varphi_{n}$
are  Klein--Gordon fields on de Sitter brane in the Euclidean
vacuum state, namely their ambient de Sitter  v.e.v. in the
$d+1$--dimensional Euclidean vacuum is given by \bea &&
W_{\lambda,\epsilon;\lambda',\epsilon'}(y,y')= \langle
\Omega|\Varphi_{q,\epsilon} (y)\Varphi_{q',\epsilon'}(y')|
\Omega\rangle  = \delta(\l-\l')\delta_{\epsilon \epsilon'}
W_\l^{(E,d)}(y,y')\cr &&W_{n;n'}(y,y')= \langle
\Omega|\Varphi_{n} (y)\Varphi_{n'}(y')| \Omega\rangle  =
\delta_{nn'} W_{\l_n}^{(E,d)}(y,y'). \eea
All other correlators vanish identically.} \\
{Here $W_\l^{(E,d)}$ is the Euclidean
two-point function in $d$ dimensions (see Eq. \ref{wbm}) 
with square mass $\lambda$.
}
By inverting now the completeness relations for the fields
 $$\PHI(X) =
\sum_n \theta_n(x)\Varphi_n(y) +\sum_\epsilon \int_\R {\rm d}q\,
\theta_{q,\epsilon} (x)\Varphi_{q,\epsilon}(y)$$
we obtain the following decomposition of the Euclidean de Sitter 
two--point function in terms of lower dimensional ones;
this is quite a nontrivial relation between 
Legendre functions in different dimensions:
\bea
 && W^{(E,d+1)}_M(X,X')=
\sum_{n=0}^{[\rho]} \theta_n(x)\theta_n(x')
W^{(E,d)}_{\lambda_n}(y,y') +\cr
&&\hspace{2truecm} + \sum_\epsilon \int_{(d-1)^2/4}^\infty
 \!\!\!\!\!\!\! {\rm d}\lambda \,\,\theta_{(\lambda-
{(d-1)^2}/4)^{1/2},\epsilon}(x)\theta_{(\lambda-
{(d-1)^2}/4)^{1/2},\epsilon} (x')
W^{(E,d)}_{\lambda}(y,y') = \cr
&&=
\sum_{n=0}^{[\rho]} \overline{P^{-\rho+n}_\rho(\cos(x')
+i\epsilon)}P^{-\rho+n}_\rho(\cos(x)
+i\epsilon) \frac{(\rho-n)
\Gamma(1+2\rho-n)} {n!\le(\sin(x)\sin(x')\ri)^{\frac
{d-1}2}}  W^{(E,d)}_{\l_n}(y,y')+\cr \vrul
& &+ \int _\R \frac { {\rm d}q \sinh(\pi q)q}{2\pi^2
\le(\sin(x)\sin(x')\ri)^{\frac {d-1}2}}
\left|\Gamma(1+\rho-iq)\Gamma(-\rho-iq)
\right|^2\cdot\cr\vrul
&& \cdot\left[e^{\pi
q}\overline{P^{iq}_\rho(\cos(x')+i\epsilon)}P^{iq}_\rho(\cos(x)+i\epsilon)
+ e^{-\pi q}
\overline{P^{iq}_\rho(-\cos(x') -i\epsilon)}
P^{iq}_\rho(-\cos(x) -i\epsilon) \right]\cdot \cr\vrul
&&\hspace{8truecm} \cdot W^{(E,d)}_{q^2+\frac{(d-1)^2}4}(y,y').
\label{expans2}
\eea
On a fixed  de Sitter brane $x=x'$ we get a
K\"allen--Lehmann type decomposition of the correlator of the bulk quantum
field, with a measure given by
\bea
& &\mu(q,x)=\sum_0^{[\rho]} \left|P^{-\rho+n}_\rho(\cos(x)
+i\epsilon)\right|^2\frac{(\rho-n)
\Gamma(1+2\rho-n)} {n!\sin^{d-1}(x)}\delta(q-(\rho-n))+\cr
& &+ \frac { \sinh(\pi q)q}{2\pi^2 \sin^{d-1}(x)}
\left|\Gamma(1+\rho-iq)\Gamma(-\rho-iq)
\right|^2 \left[e^{\pi
q}|P^{iq}_\rho(\cos(x)+i\epsilon)|^2\ri. +\cr
&& +\le. e^{-\pi q}
|P^{iq}_\rho(-\cos(x) -i\epsilon)|^2\right]\ .
\eea
In all these formul\ae\ the discrete contribution vanishes whenever
$M^2\geq\le(\frac {d-1}2\ri)^2$.\par
We remark that the formula of this decomposition matches the one in
\cite{moschellaschaeffer} which was obtained by the 
completely different method of Laplace-type transform.
\subsection{Decomposition of Minkowski states into uniformly
accelerated world--lines (Unruh effect)}
\label{app2}
In this section we revisit the Unruh effect; the general
framework is the same as in the previous examples, except that
now the codimension of the leaves $\Y$ is maximal (i.e. $d$,
where the dimension of the ambient manifold is $d+1$). What is
new in the present approach to this old model, is that we obtain
a closed formula for the decomposition of the ambient QFT into a
collection of harmonic oscillators which oscillate in the proper
time of the accelerated observer and not in the  time of an
inertial observer. Now the ambient manifold  is the {\em wedge}
$\M=\{X\in {\mathbb M}^{d+1}:\ |X^0|<X^d\}$
 of a Minkowskian spacetime
 ${\mathbb M}^{d+1}$ and  $\Y$ is
the unidimensional world-line of an accelerated observer.
In this case, the field $\PHI$ will be  reduced to a set of
harmonic oscillators.\par
An uniformly accelerated world-line
is conveniently parametrized by
$$(\xi\sinh \tau,\vec x,\xi\cosh \tau)$$
where $\vec x$ are the remaining $d-1$ coordinates in  Minkowski space.
In terms of these coordinates the wedge acquires the structure
of warped product
of a $d$--dimensional
Riemannian half space $\X=\R^d_+$  with a $1$--dimensional timelike line
$\Y$, with warping function $\omega(\xi,\vec x) = \xi$:
\be
ds^2=\xi^2 d\tau^2-d\xi^2-\sum_1^{d-1} (dx^i)^2.
\ee
The transverse problem  is
\bes
\xi^2(\tilde\triangle_\X+M^2) \theta(\xi,\vec x) =\xi^2\left[
-\pa_\xi^2 -\sum_1^{d-1}\left(\frac\pa{\pa
x^i}\right)^2-\frac 1 \xi \pa_\xi\ +M^2\right] \theta (\xi,\vec x)=\lambda\theta\ ,
\ees
and the corresponding Hilbert  product
\bes
 (\varphi,\psi)=\int_{\R^d_+} \frac {{\rm d}\xi}\xi\le(\prod_1^{d-1} dx^i\ri)\,
\overline\varphi(\xi,\vec x)\, \psi(\xi,\vec x).
\ees
A straightforward computation produces the following generalized
orthonormal eigenfunctions
\bea
 \theta_\lambda(\xi,\vec x)&=& \theta_{m,\vec p,\pm}(\xi,\vec x) =  \frac{
\sqrt{2\, \sinh(\pi  m)}}\pi K_{i m}\le(\xi\sqrt{M^2+\vec p^2}\ri)
\frac {1}{(2\pi)^{\frac {d-1}2}} \le\{ \matrix{\cos(\vec p\cdot\vec x)\cr
\sin(\vec p \cdot\vec x)}\ri\} =\cr\vrul
&&=
  N_m  K_{i m}\le(\xi\sqrt{M^2+\vec p^2}\ri)
\le\{ \matrix{\cos(\vec p\cdot\vec x)\cr
\sin(\vec p \cdot \vec x)}\ri\} ,\\
\eea
where the $\pm$ subscript selects among $\cos(\vec p \cdot \vec x)$
and $\sin(\vec p \cdot \vec x)$.
In this case the eigenvalue $\lambda=m^2$ has a $\R^{d-1}$
degeneracy.
Again, we introduce the quantum fields
\bes
\Varphi_{\lambda, \vec p,\pm} (\tau) =\int_0^\infty \frac {{\rm d}\xi}\xi
\int_{\R^{d-2}} \!\!\!\!\! \!\!\!\!\!  {\rm d}\vec
 x\,\, \theta_{\lambda,\vec p,\pm} (\xi,\vec x) \PHI(\tau,
\xi,\vec x)\ .
\ees

Let now $W(X,X')$ be the usual Wightman two--point function for the
quantum field $\PHI(X)$ as given by Eq. (\ref{01}):
we can directly compute the correlators $W_{m,\vec p,\epsilon;m'\vec
p',\epsilon'}(\tau,\tau')$
of the fields
$\Varphi_{\lambda, \vec p,\pm} (\tau)$
and
show that they are diagonal in $m$, $\vec p$ and the discrete index
$\epsilon\in \{+,-\}$.
Indeed
\beas
&& W_{m,\vec p,\epsilon ;m',\vec p',\epsilon'}(\tau,\tau')=
\int_{\R^d_+} \frac {{\rm d}\xi}\xi d\vec x\int_{\R^d_+} \frac {{\rm d}\xi'}{\xi'}
d\vec x'
\theta_{m,\vec p, \epsilon}(\xi,\vec x)\theta_{m',\vec p',\epsilon'}(\xi',\vec x')W(X,X')
=\cr\vrul
&&= \int_{\R^d_+} \frac {{\rm d}\xi}\xi d\vec x\int_{\R^d_+} \frac
{{\rm d}\xi'}{\xi'}
d\vec x'
N_m  K_{i m}\le(\xi\sqrt{M^2+\vec p^2}\ri) \le\{ \matrix{\cos(\vec p\cdot\vec x)\cr
\sin(\vec p \cdot \vec x)}\ri\}\cdot\cr
&&\hspace{1truecm} \cdot
N_{m'}  K_{i m'}\le(\xi'\sqrt{M^2+\vec p'^2}\ri)
 \le\{ \matrix{\cos(\vec p'\cdot\vec x)\cr
\sin(\vec p' \cdot \vec x)}\ri\}
\frac 1{(2\pi)^{d+1}}
\cr\vrul
&&\hspace {1cm} \int_{\R^{d+1}} \frac{{\rm d}^{d+1}P}{(2\pi)^d}
\,\delta(P^2-M^2)
\Theta(P_0) e^{iP\cdot (X-X')}=\cr\vrul
&&= N_m N_{m'}\int_{\R_+}\frac {{\rm d}\xi}\xi \int_{\R_+}\frac {{\rm d}\xi'}{\xi'}
K_{i m}\le(\xi\sqrt{M^2+\vec p^2}\ri)K_{i m'}\le(\xi'\sqrt{M^2+\vec
p'^2}\ri)\delta(\vec p-\vec p')\delta_{\epsilon,\epsilon'}\cr\vrul
&& \int_{\R^2}
\frac{{\rm d}P_0{\rm d}P_1}{2\pi}
\delta(P_0^2\! -\!P_1^2\!\! -\vec p^2\!\! -M^2) \Theta(P_0) {\rm e}^{
iP_0\le(\xi\sinh(\tau)-\xi'\sinh(\tau')\ri) +
iP_1\le(\xi\cosh(\tau)-\xi'\cosh(\tau')\ri)}.
\eeas
The remaining integration is a special case of Formula (\ref{expr})
 with the substitutions $M^2\mapsto M^2+\vec p^2$,
$\nu\mapsto m$, $d=1$. This finally gives
\bes
W_{m,\vec p,\epsilon ;m',\vec p',\epsilon'}(\tau,\tau')= \delta(\vec
p-\vec p')\delta_{\epsilon,\epsilon'}
\delta(m^2-{m'}^2) \frac {\cos \le(m(\tau-\tau') + i\pi m\ri)}
{2 m \sinh(\pi m)}.
\ees
This expression  is the Wightman function of an harmonic oscillator  in a
thermal state at an inverse temperature $\beta$ (in the
Heisenberg picture):
indeed, the quantum Klein--Gordon field on a one--dimensional
spacetime corresponds to a single quantum harmonic
oscillator in the Heisenberg picture where the mass represents
the spring constant.
The  thermal time correlation function of the
position operator
at inverse temperature $\beta$ for such  oscillator  is given by:
\be
W(t,t')=\frac {\cos(\omega (t-t' +i\beta/2))}{2\omega \sinh(\omega\beta/2)}
\label{qo}\ee
which is precisely the expression derived above with $\beta = 2\pi$.\\
Using the completeness of the modes $\theta$ we can express the vacuum
two--point function of the field $\PHI$ in terms of the two--point
functions of the thermal oscillators as in
\bes
W(X,X') = \sum_{\epsilon = +,-} \int_{\R^{d-1}} \!\!\!\!\!\!\! {\rm
d}\vec p \int_0^\infty {\rm d}(m^2)
\theta_{m,\vec p,\epsilon} (\xi,\vec x)\theta_{m,\vec p,\epsilon}
(\xi',\vec x')  \frac {\cos \le(m(\tau-\tau') + i\pi m\ri)}
{2 m \sinh(\pi m)}
\ees
In this case we know that if the state of the ambient field $\Phi(X)$ is the
usual vacuum one,  the quantum theory obtained from the ambient
space one is {\em thermal} at the Unruh inverse temperature
$\beta_U={2\pi} $. 
The decomposition
\be
\bra{\Omega}\Phi(\tau,\xi,\vec x)\Phi( \tau', \xi,\vec x)\ket{\Omega}= \int
\mu((\xi,\vec x),m) <\Varphi_m(\tau)\Varphi_m(\tau')>_{\beta_U}
\ee
defines correlation functions $<\varphi_m(\tau)\varphi_m(\tau')>_{\beta_U}$
of a {\em thermal state} of the quantum harmonic oscillator given by $\frac
{d^2}{d\tau^2} \varphi_m(\tau) +m^2\varphi^2(\tau)=0$.
Note that along each uniformly accelerated world-line, specified by the 
parameters $\xi$ and $\vec x$, the corresponding proper-time is equal to 
$\xi \tau$ ($\xi $ being the value of the Tolman factor),  
so that the temperature ``really felt by the corresponding observer'' 
on this world-line   
is equal to $\frac {1} {2\pi \xi}$.  

\subsection{AdS states in terms of Minkowski states}
\label{app3}
This last example concerns the states of a Klein--Gordon field theory
on  the AdS spacetime foliated by flat Minkowski spacetimes of
codimension one: this decomposition has been used in \cite{BBMS:1999}
in application to the AdS--CFT correspondence and it will be just briefly
reported.\par
  This example lies somewhat outside of the picture we
have drawn in the general part because the AdS spacetime is not
globally hyperbolic. Nevertheless we can prove directly that a
completely analogous decomposition of the Klein--Gordon field can be
achieved.\\
To set the notation, in the spirit of Section
\ref{dS2dS}
we consider  the vector space ${\mathbb R}^{d+2}$ equipped with the
following pseudo-scalar product:
\begin{equation}
X\cdot X'  = {X^0} {X'}^0 - {X^1} {X'}^1 - \cdots - {X^d}{X'}^d
+ X^{d+1}  {X'}^{d+1}\ .
\label{ambientmetric}
\end{equation}
The $(d+1)$-dimensional AdS universe can then be identified with the
 quadric
\begin{equation}
AdS_{d+1} = \{ X \in \mathbb R^{d+2},\;\; {X^2}=R^2\},
\end{equation}
where  $X^2= {  X} \cdot{X}$, endowed with the induced metric
\begin{equation}
{\mathrm d}s^2_{AdS} = \left.\left(d{{X}^0}^{\,2}-d{{X}^1}^{\,2}
- \cdots + d{{X}^{d+1}}^{\,2}\right) \right|_{AdS_{d+1}}.
\label{metric}\end{equation}
The AdS relativity group  is   $G =SO_0(2,d )$, that is
the connected  component of the identity of
the pseudo-orthogonal group $SO(2,d )$.
Two events  $X$, $ X'$ of $AdS_{d+1}$
are space--like separated if $(X- X')^2<0$,
i.e. if  $X\cdot  X'>R^2$.
In the following we will put for
notational simplicity $R=1$.\\[5pt]
We consider an open subset of AdS given by the inequality in the
ambient space $\Pi\equiv\{X^d + X^{d+1}>0 \}$: this is ``half'' the spacetime.
In the {\sl  ``horocyclic parametrization'' $X = X(x,y)$}, there appears
a {\em structure of warped product}: this set of coordinates covers  $\Pi$
and is obtained by intersecting $AdS_{d+1}$ with
the hyperplanes $ \{X^{d} + X^{d+1} = e^ {x} =\frac{1}{s}\}$
each slice $\Pi_v$ (or ``horosphere'') being an hyperbolic paraboloid:
\begin{equation}
\left\{\begin{tabular}{lclcll}
 $X^{\mu} $&=& $e^{ {x}} y^\mu  $ & =& $\frac{ 1}{s}y^\mu  $& $
 { \mu=0,1,...,d-1}$\label{AdSprod}\\
 $X^{d} $&=& $\sinh  {x} + \frac 12 e^{ {x}} y^2 $  & =& $\frac{1-s^2}{2s} +
\frac {1}{2 s} y^2$ & $y^2 ={ y^0}^2- {y^1}^2- \cdots -{y^{d-1}}^2$  \cr
 $X^{d+1}$&=&$ \cosh  {x} - \frac 12 e^{ {x}} y^2$ &  =& $\frac{1+s^2}{2s} -
\frac {1}{2 s} y^2$&
\label{coordinates}
\end{tabular}\right.
\end{equation}
In each slice $\Pi_v$, $y^0,...,y^{d-1}$ can be seen as coordinates of
an event of a $d$-dimensional
Minkowski spacetime ${\mathbb M}^{d}$ with metric $ds^2_{{M}}= d{{y^0}}^{2}-
d{{y^1}}^{2} - \ldots -d{{y^{d-1}}}^{2}$
(here and in the following where it appears, an index {\em {\small M}} stands for Minkowski).
This explains why  the  horocyclic coordinates $(x,y)$ of the
parametrization   (\ref{coordinates})
are also called  Poincar\'e coordinates.
The scalar product (\ref{AdSprod}) and the AdS metric can then be rewritten as follows:
\begin{eqnarray}
&& X\cdot X' = \cosh( {x}- {x} ')  - \frac 12 e^{ {x}+ {x}'} \le(y-y'\ri)^2,
\label{7}\\
&& {\mathrm d}\sigma^2_{AdS} = e^{2 {x}} {\mathrm d}\sigma^2_{{M}}-{\mathrm d} {x}^2
= \frac{1}{s^2}({{\mathrm d}\sigma^2_{{M}}-{\mathrm d} s^2}). \label{metric1}
\end{eqnarray}
Eq. (\ref{metric1}) exhibits the region $\Pi$ of $AdS_{d+1}$ as a
warped product with warping function $\omega( {x})=e^{{x}}$
 and fibers conformal to ${\mathbb M}^{d}$.
\vskip 0.4cm
\noindent
We apply the formalism of Section \ref{warped} and obtain the spectral
problem
\begin{eqnarray}
&& e^{2 {x}} \left[\theta''( {x}) + d \, \theta'( {x}) -M^2\theta( {x})\right] =
-\lambda \theta( {x}),\label{equat}
\end{eqnarray}
to be considered in the Hilbert space
$L^2( {\mathbb R} , e^{(d-2)x} {\mathrm d} x )$,
where the differential operator defined in Eq. (\ref{equat}) is symmetric.
In the variable $s = e^{-x}$
already introduced  in Eq. (\ref{coordinates}) and defining
$f(s)=\theta( {x}) e^{\frac{d-1}{2}  {x}} $ Eq. (\ref{equat})
is turned into the  well-known Schr\"odinger spectral problem on the half-line
\begin{equation}
-f''(s) + \frac{M^2+\frac{d^2-1}4}{s^2} f(s)= -f''(s) +
\frac{(\nu+1/2)(\nu-1/2)}{s^2} f(s)
=\lambda f(s)\ .
\label{bessel}
\end{equation}
Following \cite{Titchmarsh:1962}, pag. 88 ff, we learn that  there are
two distinct regimes corresponding  to the two ranges $\nu \geq 1$ and
$|\nu|<1$.

When $\nu\geq 1$ the previous operator is essentially self--adjoint
and  there is only one
possible choice for the generalized eigenfunctions, namely
 \be
f_\l(s)=\frac 1 {\sqrt{2}} \,   s ^\frac 1 2 J_\nu\le(\sqrt{\lambda} \,s \ri)\ ,
\ee
where $J_\nu$ are Bessel's functions.
The completeness of these eigenfunctions gives Hankel's formula,
which expresses the
resolution of the identity in $L^2({\mathbb R} ^+, {\rm d}s)$ as follows:
\be
g(s) = \int_{0}^\infty {\rm d}\l\, f_\l(s)\int_0^\infty f_\l
(s') g(s') {\rm d} s'\ ,\ \ \forall g\in L^2({\mathbb R} ^+, {\rm d} s))\ .
\ee
When $0\leq \nu< 1$ both  solutions $s^{1/2}J_\nu(\sqrt{\lambda}s)$ and
$s^{1/2}J_{-\nu}(\sqrt{\lambda} s) $  are square integrable in the
neighborhood of $s=0$ and
must be taken into consideration:
we are in the so--called {\em limit circle case} at zero \cite{Titchmarsh:1962,Reed},
which implies that the
operator is not essentially self--adjoint and there exists a $S^1$
ambiguity in the self--adjoint extensions we can perform.
The freedom  is exactly in the choice of the boundary conditions at
$s=0$ (corresponding to the   boundary of AdS).\\
Now  we have a one--parameter family of eigenfunctions:
 \be
f^{(\K)}_\l (s) \equiv \sqrt{\frac {s} {2}}{{ \le(\K^2 - 2\K\l^\nu\cos(\pi\nu) +
\l^{2\nu} \ri)^{-\frac 1 2}}}
\le[\K\, J_{\nu}(\sqrt{\l}\, s) -\l^{\nu}  J_{-\nu}(\sqrt{\l}\, s) \ri]\
,
\ee
to which we must add one  bound state when  $\K>0$:
\be
f^{(\K)}_{\rm bound}  (s) \equiv \sqrt{ 2\K^{\frac 1 \nu}\frac
{\sin{\pi\nu}}{\pi\nu} } s^\frac 1 2  K_{\nu} (\K^{\frac 1 {2\nu}} s)\ .
 \ee
The possible choices of the parameter
$\K$ do correspond to different self--adjoint extensions of the  differential
operator (\ref{bessel}). To each such extension there is associated a
domain ${\mathfrak D}^{(\K)}$ also depending on the parameter $\K$ \cite{Reed}.
To construct ${\mathfrak D}^{(\K)}$ consider the one dimensional subspaces
$H_\pm$ spanned by the  eigenfunctions solving
Eq. (\ref{bessel}) with eigenvalues $\pm i$:
\be
f_{\pm} (s) \equiv \sqrt{s} K_\nu (e^{\pm \frac {i\pi}4} s)\ ;
\ee
both these functions are square-integrable when $0\leq \nu<1$.
Each extension
is in one--to--one correspondence with partial isometries
$U:H_+\mapsto H_-$, namely --in this case-- with elements of
$U(1)\simeq S^1$. The domain of the extension is obtained by
adjoining to the original domain of symmetry the subspace $\le( {\rm
id}_{H_+}+ U\ri) H_+$: here it means that we have to add the span of
the $L^2$ element
\bes
f_\alpha(s)\equiv f_{+}(s) + e^{i\alpha} f_{-}(s)\ .
\ees
which has in our case the asymptotics
\bea
f_\alpha(s) \simeq \frac {\pi}{2\sin(\pi\nu)} \le[ \frac {2^\nu\le(
e^{-\frac {i\pi\nu} 4} + e^{i\alpha + \frac {i\pi\nu}4}  \ri)}
{\Gamma(1-\nu)} s^{-\nu} -  \frac {2^{-\nu}\le(
e^{\frac {i\pi\nu} 4} + e^{i\alpha - \frac {i\pi\nu}4}  \ri)}
{\Gamma(1+\nu)} s^{\nu}
\ri]\ .
\label{asymext}
\eea
The generalized eigenfunctions of the operator (\ref{bessel})
corresponding to  a specific extension have the following asymptotics
\be
f^{(\K)}_\l(s) \simeq  2^{-\frac 1 2}
  s^{\frac 1 2}  \le(\K^2 - 2\K\l^\nu\cos(\pi\nu) +
\l^{2\nu} \ri)^{-\frac 1 2}  {\l}^{\frac \nu 2}
\le[\K\, \frac {2^{-\nu}  s^\nu} {\Gamma(1+\nu) }
 - \frac
{2^{\nu} s^{-\nu} }{\Gamma(1-\nu)} \ri]\ .
\label{asymmext}
\ee
As usual these functions do not belong to $L^2({\mathbb R} ^+, {\rm d} s)$ but any
wave--packet does; moreover any such wave packet has this asymptotics.
This allows us to find which parameter $\K$ corresponds to which
unitary operator $e^{i\alpha}:H_+\mapsto H_-$, i.e. to a specific
self--adjoint extension. Indeed, by matching the asymptotics in
Eqs. (\ref{asymext}) with that in Eq. (\ref{asymmext}) we obtain
\bes
\K = \frac{\cos\le(\frac \alpha 2-\frac {\pi\nu}4 \ri)}{\cos\le( \frac
\alpha 2+ \frac {\pi\nu}4\ri)}\ .
\ees

\vskip 0.5cm 
We consider now a very specific QFT on the AdS spacetime: this QFT is
a generalized free field theory which satisfies certain analyticity
properties \cite{Bros:1999}. It depends on the single
(complexified) invariant $\zeta = Z\cdot Z' = \cosh(x-x') - \frac 1 2
e^{x+x'}\le(z-z'\ri)^2$, where now $z$ (respectively, $z'$) belongs
 to the complexified
Minkowski space and its imaginary part lies in the interior of the
future (resp.  past) light cone.\\
Such a QFT is characterized by the
$SO(2,d)$--invariant two--point function given by
\begin{equation}
W^{d+1}_{\nu}(Z,Z') = w_\nu(\zeta) =
\frac {e^{-i\pi\frac {d-1}2}}{(2\pi)^{\frac{d+1}2}}
(\zeta^2-1)^{-\frac {d-1}4} Q^{\frac {d-1}2}_{\nu-\frac 1 2}(\zeta).
\label{kgtp}
\end{equation}
The analyticity domains advocated in \cite{Bros:1999}  are such that the complex
variable $\zeta$ belongs to the complex plane cut along the segment
from $-1$ to $1$ (the ``causal cut''). The analogous invariant
variable in the Minkowskian case is $\delta= -(z-z')^2$ and the
causal cut in this case is the negative real axis: the ``Euclidean
regime'' corresponds to positive real values of $\delta$.
We can now show   by direct computation that the
two--point function (\ref{kgtp})
in $AdS_{d+1}$ in the whole range $\nu\in (-1,\infty)$
can be decomposed as follows:
\bea
&&
W^{d+1}_{\nu}(Z( {x},z),Z'( {x'},z'))= \int_0^\infty {\rm d}\l
\theta_\l( {x})\theta_\l( {x}') W^{{{{M}}},d}_\lambda(z,z')\ , \
\ \nu\in[1,\infty) \cr
&&
W^{d+1}_{\nu}(Z( {x},z),Z'( {x'},z')) =\int_0^\infty\!\!\!\!\! {\rm d}\l
\theta_\l^{(\infty)}( {x})\theta_\l^{(\infty)} ( {x}')
W^{{{{M}}},d}_\lambda(z,z'), \
 \nu\in [0,1) \cr
&&
W^{d+1}_{\nu}(Z( {x},z),Z'( {x'},z')) =\int_0^\infty {\rm d}\l
\theta_\l^{(0)}( {x})\theta_\l^{(0)}( {v}') W^{{{{M}}},d}_\lambda(z,z'), \
 \nu\in (-1,0),\cr
&&\
\label{deco}
\eea
where $W^{{{{M}}},d}_\lambda(z,z')$ is the usual  two--point function
for a Klein--Gordon field on $\mathbb M^d$ of square mass $\lambda$ in the
Wightman vacuum:
\bea
&&W^{{{{M}}},d}_\lambda (z,z') \equiv \int \frac{{\rm d}^d p}{(2\pi)^{d-1}}
\delta(p^2-\l)\Theta(p_0) e^{-ip\cdot(z-z')} = \cr
&&\hspace {2cm}=
(2\pi)^{-\frac d 2} \le(\frac {\delta }{\sqrt{\lambda}} \ri)^{\frac
{2-d}2} K_{\frac {d-2}2}\le(\sqrt{\lambda} \delta  \ri)\ ;\qquad \delta\equiv
-(z-z')^2\ .
\label{tpmink}
\eea
In Eqs. (\ref{deco})  the functions $\theta_\l^{(\infty)}$  and the
$\theta_\l^{(0)}$ belong to the domains of self--adjointness corresponding to
the values  $\K=\infty$ and $\K=0$ respectively. They  explicitly read
\bea
&& \theta_\l^{(\infty)} ( {x}) = \frac 1{\sqrt {2}} e^{-\frac d 2  {x}
} J_\nu(\sqrt{\l} e^{- {x}}) \label{modes>1}\\
&& \theta_\l^{(0)} ( {x}) = \frac 1{\sqrt {2}} e^{-\frac d 2  {x}
} J_{-|\nu|}(\sqrt{\l} e^{- {x}})\ .
\eea
The reason why we must use different self--adjoint extensions is that
$W^{d+1}_\nu( Z({x},z) ,$ $ Z({x}',z') )$, as a function of
$ {x}$ (or $ {x}'$) belongs to ${\mathfrak D}^{(\infty)}$ when
$\nu\in[0,1)$ while it belongs  to ${\mathfrak D}^{(0)}$ when
$\nu\in (-1,0)$:
this can be  proved directly by studying the asymptotics.\par
The three Eqs. (\ref{deco}) are thus summarized into the following
formula valid for the whole range of parameter $\nu$:
\bea
&& W^{d+1}_{\nu}(Z( {x},z),Z'( {x'},z'))=\cr
&& =(2\pi)^{-\frac d 2} ({s}\,{s}')^{\frac d 2}
\int_0^\infty \frac{{\rm d}\lambda} 2 \, \lambda^{\frac {d-2} 4} J_\nu(\sqrt{\lambda} \,{s})
J_\nu (\sqrt{\lambda}\,{s}')  \delta ^{\frac {2-d}2 }\, K_{\frac
{d-2}2} (\sqrt{\lambda}\,\delta)\ ,
\label{MAdS}
\eea
with, again,  $s=e^{-x}$.\\
The proof is an application of  formula (12) pag. 64 in
\cite{Bateman2}, which is the Hankel's
transform of the product of two Bessel's functions (we simply adapt the notation)
\beas
&& \int_0^\infty {\rm d}m\, m^{\mu+\frac 1 2 } J_\nu(m\, s) J_\nu(m\,s') K_\mu
(m\,\delta ) (m\, s')^\frac 1 2 =\cr
&&=  \frac {\delta^\mu s^{-\mu -1} {s'}
^{-\mu-\frac 1 2 } e^{-(\mu+\frac 1 2 )i \pi} } {\sqrt {2\pi}} (\zeta^2-1)^{-\frac
\mu 2-\frac 14 } Q^{\mu+\frac 1 2}_{\nu-\frac 12 }(\zeta)\ ,\\
&& \Re (\nu)>-1\, ,\qquad \Re(\mu+\nu)>-1\ ,
\eeas
where $\displaystyle{ \zeta=\frac {s^2+{s'}^2+\delta^2}{2s\,s'}}$. Here
we implicitly perform the ``Wick rotation'' to the Euclidean section
where $\delta>0$ and hence $\zeta= \cosh(x-x')+\frac 1 2
e^{x+x'}\delta >1$.\par
Since the modes $\theta_\lambda$ form a orthonormal basis in the
Hilbert space,
Eq. (\ref{deco}) can also be inverted and we obtain
the Minkowski Klein--Gordon two-point function on the slice  $\Pi_v$
by smearing $W_\nu$ against the eigenfunctions $\theta_\lambda$.
For instance, when $\nu>1$ this corresponds to the introduction of the fields
$\Varphi_{\lambda}(y)$ on the Minkowskian slice
$\Pi_v$ obtained by smearing the AdS Klein--Gordon
field $\PHI$  with  the complete
set of    modes  (\ref{modes>1}):
\be
\Varphi_{\lambda } (y)= \int_{-\infty}^{\infty}
\PHI(X(x,y)) \theta_{\lambda} (x) e^{(d-2)x} {\rm d}x .
\ee
It can be shown  that the field $\Varphi_{\lambda } (y)$
is a canonical Minkowskian
Klein--Gordon field in the Wightman vacuum
state.
In precise terms, we have that the AdS vacuum expectation value
of $\Varphi_{\lambda } (y)$ is given by
\be
W_{\lambda,\lambda'}(y,y')\equiv
\langle \Omega|\Varphi_\l(y)\Varphi_{\l'}(y')|\Omega\rangle = \delta(\l-\l')
W^{{M},d}_\lambda(y,y').
\label{restrizioneAdS}
\ee
 In particular,  the fields $\Varphi_\l$ have zero correlation
(and hence commute) for different values of the square mass $\l$.\par\vskip 4pt
As a specification of the Eqs. (\ref{deco}), when restricting the AdS
Klein--Gordon field $\PHI$ to a fixed slice $\Pi_v$ (the
$d$-dimensional brane) we obtain the
following explicit formula for the K\"allen--Lehmann decomposition of
the field in the Minkowskian slice
\be
W^{d+1}_{\nu}(X( {x},y),X'( {x},y')) =\int_0^\infty \frac {{\rm d}\l}2
 e^{-d {x}
} \le(J_\nu(\sqrt{\l} e^{- {x}})\ri)^2 W^{{{{M}}},d}_\lambda(y,y').
\ee
This formula is telling us that a free field $\Phi$ propagating in the
ambient gravitational background will be seen on the $d$-dimensional
brane as a superposition of fields with a continuous spectrum of
masses but different relative weight given by
\be
{\rm d}\mu(\l,x) =
\frac {{\rm d}\l}2
 e^{-d {x}
} \le(J_\nu(\sqrt{\l} e^{- {x}})\ri)^2\ .
\ee
\vskip 3pt
The results of this section can be used to  construct other
two--point functions\\ $W^{d+1,(\K)}_\nu (X(x,y),X(x',y'))$
 for a  Klein--Gordon field  on AdS by using the other self--adjoint
extensions: however it is not guaranteed that such  $W^{d+1,(\K)}_\nu$ can be
extended to the other half of AdS since the definition uses the set of
coordinates defined only on one half. Moreover one should prove (or disprove)
the AdS invariance and analyticity properties of such states.
We will not go any further in this direction in this paper.
\section{Conclusions}
We have considered  a particular foliation of a Lorentzian manifold by
means of Lorentzian submanifolds over a Riemannian base: such foliation also gives
a particular orthogonal splitting of the metric tensor.
In this context we have considered a quantum field over the total
manifold and decomposed it into a bunch of {\em longitudinal quantum
fields} $\Varphi_\lambda$  and {\em transversal classical modes}
$\theta_\lambda$.\\
Such decomposition allows us to pick up a specific member of the bunch
by a smearing against those transversal classical modes.\par
This technique has been then successfully applied to the case of
Minkowski, foliated by de Sitter $d$-branes or by accelerated world-lines;\\
to the case of de Sitter, foliated by lower dimensional de Sitter
branes;\\
to Anti de Sitter, foliated by Minkowskian branes.\par
In all these cases the  distinguished analyticity properties of the two--point
function in the ambient manifold appear to survive this operation of
picking out a specific field, giving a QFT on the leaf with those analyticity
properties which are advocated independently for the geometry of the brane itself.\\
Since the analytic structure of the two--point function is equivalent
to the spectral structure of the Hamiltonian of the theory, this
procedure can be regarded as a method for enforcing certain spectral
properties on a manifold by embedding it into another manifold where
the spectral properties are easier to formulate (this is the case
de Sitter $\hookrightarrow$ Minkowski).\\
Or else we can construct a QFT with certain spectral properties in the
ambient manifold by means of the spectral properties of the QFT in the
brane (this is the case Minkowski $\hookrightarrow$ AdS).\par
We also point out that, in more geometrical terms, to some extent
what we have done in the examples is decomposing a certain irreducible unitary
representation of the invariance group of the ambient manifold into
irreducible unitary representations of a certain subgroup which is the
invariance group of a submanifold. This might turn out to be of
utility in application to representation theory and special functions:
indeed some of the relations (e.g. Eqs.(\ref{expans},\ref{expans2}))
 that we have found, relating the two-point
functions of the ambient manifold and those of the submanifold are integral
representations which are not to be found in the more mathematically
oriented literature.\par
This decomposition has been made here only for the {\em
warped--product} manifolds for practical computational issues but
nevertheless the idea of inheriting spectral properties from an
ambient manifold could be extended to other cases, most importantly
the Schwarzschild geometry \cite{bertola,deser}.\par
Potentially this perspective is the more appealing the harder is the
problem of consistently formulating a spectral property in curved
backgrounds.\par
Additionally, the recent topic raised in \cite{Randall:1999,
Randall:2000} allows a direct application of this method to general
warped $d$-branes in various gravitational backgrounds.\\
It is our intention to pursue this direction in further publications.

\end{document}